\documentclass[preprint2]{aastex}
\usepackage{graphics}
\usepackage{psfig}
\usepackage{latexsym,amssymb}

\shortauthors{F. Fusi Pecci et al.}
\shorttitle{Massive young clusters in the disc of M31}
\received{2004 November 5}
\begin{document}
\title{Massive young clusters in the disc of M31}

\author{F. Fusi Pecci, M. Bellazzini, A. Buzzoni,
	   E. De Simone, L. Federici}
\affil{INAF - Osservatorio Astronomico di Bologna,}
\affil{Via Ranzani 1, 40127 Bologna, Italy}
\email{flavio.fusipecci, michele.bellazzini,
	     alberto.buzzoni, luciana.federici@bo.astro.it}
\and
\author{S. Galleti\altaffilmark{1}}
\affil{Dipartimento di Astronomia, Universit\'a di Bologna,}
\affil{Via Ranzani 1, 40127 Bologna, Italy}
\email{silvia.galleti2@unibo.it}
\altaffiltext{1}{INAF - Osservatorio Astronomico di Bologna, Via Ranzani 1,
   40127 Bologna, Italy}
   
\begin{abstract}
   
We have studied the properties of a sample of 67 very blue and likely  young
massive clusters in M31 extracted from the Bologna Revised Catalog of globular
clusters, selected according to their color [$(B-V)_o \le 0.45$] and/or to the strength 
of their $H\beta$ spectral index ($H\beta \ge 3.5$~\AA).  
Their  existence in M31 has been noted by several authors in the past; we show
here that these Blue Luminous Compact Clusters (BLCCs) are a significant 
fraction ($\gtrsim 15$\%)  of the whole globular cluster system of M31. 
Compared to the global properties of the M31 globular cluster system, they
appear to  be intrinsically fainter, morphologically less concentrated, and
with a shallower Balmer jump and enhanced $H\beta$ absorption in their spectra.

Empirical comparison with integrated properties of clusters with known  age as
well as with theoretical SSP models consistently  indicate that their typical
age is less than $\sim 2$~Gyr, while they probably are not so metal-poor as
deduced if considered to be old. Either selecting BLCCs by their $(B-V)_o$
colors or by the strength of their $H\beta$ index  the cluster sample turns out
to be  distributed onto the outskirts of M31 disc, sharing the kinematical
properties of the  thin, rapidly rotating disc component.

If confirmed to be young and not metal-poor, these clusters indicate the
occurrence of a significant recent star formation in the thin disc of M31,
although they do not set constraints  on the  epoch of its early formation.

\end{abstract}

   \maketitle 


\section{Introduction}

Globular clusters (GCs) are ubiquitous stellar systems living in any kind of
galaxies, from dwarf to giants, from the earliest to the latest types. Their
integrated properties carry crucial information on the physical characteristics
of their host galaxy at the time of their formation. Hence, the study of
globular clusters systems is a fundamental tool to understand the evolutionary
history of the barionic component of distant galaxies (see \citealp{h01} and
references therein).

In this framework, the globular cluster system of the Andromeda galaxy  (M31)
plays a twofold role, as a natural reference to compare the Milky Way (MW)  GC
population, and as a fundamental testbed for the techniques to be applied to
systems in more distant  galaxies \citep[see][ and references
therein]{barmby,puzia,rich,bar03,silvia}. Indeed, the comparison of the GC
system of M31 and the MW has revealed both  fundamental similarities and
interesting differences whose complete understanding may have a deep impact on
our knowledge of galaxy formation and evolution \citep{h_book,sydbook,
morrison, beas, burst04}. 

\begin{figure}[!t]
\centerline{
\psfig{file=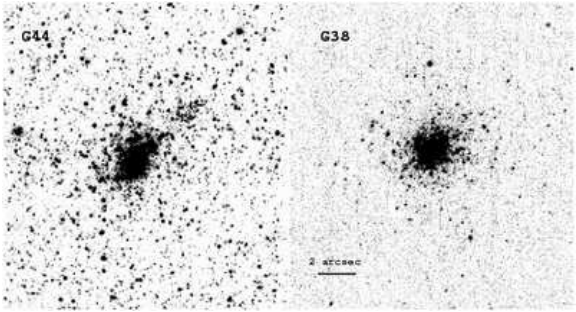,width=\hsize,clip=}}
\caption{The BLCCs G38 and G44 from the HST-WFPC2 observations by 
\citet{wh01}. Frames have been retrived from the WFPC2 associations
archive ({\tt http://archive.stsci.edu/hst/wfpc2/\protect \\
index.html)} }
\label{map}%
\end{figure}

Among the latter ones, in the present contribution we will especially  focus on
the claimed presence in M31 of stellar systems  similar to MW globulars in
luminosity and shape but with integrated colors  significantly bluer than the
bluest MW counterparts.  While some of the faintest objects may hardly be
distinguished from bright open clusters, the typical family member appears 
quite similar to classical globulars \citep[two typical
examples are shown in Fig.~\ref{map}, see][]{wh01,wh01b,h_open}. 
In Sec.~3.4 we will provide some evidences suggesting that clusters of similar age
and total luminosity may be lacking in the Milky Way.
Hereafter,  we call them ``Blue Luminous Compact Clusters'' 
(BLCCs)\footnote{We do not attach any special meaning to this newly
introduced term. It must be intended just as a convenient label to describe
their color and structural morphology to be used in
the following for sake of brevity and clarity.}

The peculiar colors of BLCCs have been first reported by  \citet{vet62}, 
\citet{syd67,syd69}, \citet{searle}, and  this class of objects then 
received growing attention
\citep{ccsc85,cow,EW88,kl91,bohlin,barmby,wh01,beas,burst04}, although a
systematic study is still lacking.

In particular, \citet{EW88} noted 14 such blue clusters,  not included in the
list of open cluster candidates by \citet{h_open}, and better consistent with a
GC morphology. Their absolute luminosities spanned the  luminosity range
$-9.5<M_{V}<-6.5$, and their positions in a two-color diagram  pointed to a
possibly young age.  For ten of these objects, \citet{kl91} provided
supplementary UBVR photometry  indicating a global luminosity around $3~10^4
\to 4~10^5$~L$_\odot$.  Based on stellar population models, their estimated age
appeared to be less than a few $10^8$~yrs, with a typical mass between $3~10^3$
and $5~10^4$~M$_\odot$. If confirmed, these values indicate that they are
higher than Galactic open clusters, but comparable to those of young, rich
globulars found in the Large Magellanic Cloud \citep{EF85,vdb91}.

\citet{boh88,bohlin}, studying the UV-colors of a sample of 49 GC candidates in
M31, listed 11 objects classified as blue clusters based on their location in
the two-color diagram, and suggested that they are probably young. On the same
line, \citet{barmby} noted that their M31 catalog of  GC candidates may be
contaminated by several young obiects with $B-V<0.55$ and they eventually
excluded 55 such objects from their analysis of old M31 clusters. 

As already stressed long ago \citep{SS72}, the integrated spectrum and color of
a cluster, especially in the blue, are influenced by the metal abundance and
the position of the main sequence turnoff stars (MSTO) (in turn, by the cluster
age), by the strength of the horizontal branch (HB), and, to a lesser degree,
by the overall luminosity function of  its composing stellar population. To
disentangle the different effects it is thus very  important to obtain the
color-magnitude distribution of cluster stellar populations. In this regard,
\citet{wh01} obtained deep HST photometry of individual  stars and c-m diagrams
for four of these BLCCs leading to estimate ages in the range  60-160~Myr and
metallicity from solar to 2/5 solar. This clearly supports the evidence that
the exceedingly blue integrated colors of BLCCs are direct consequence of
their  remarkably young age. 

\citet{beas} reached similar conclusions for eight BLCCs  by comparing high-quality
low-resolution spectra  of a sample of M31 clusters with similar data for MW
and  Magellanic Clouds globulars. \citet{burst04} reported a global sample of
19 BLCCs  in M31, including 13 ``young'' objects from the \citet{barmby} 
survey,\footnote{\citet{barmby} classified these clusters as possibly young
because of the strong Balmer absorption lines observed in their high-resolution
spectra.} most of them sharing the kinematical properties of a wider sample of 
``red'' clusters belonging to the cold thin disc detected by \citet{morrison}.

In summary, various observations suggest that M31 may have many  more young GCs
than the MW, and the latest results on the claimed existence of a thin disc
subsystem of GCs in M31, quite large in number and covering a very wide range
in metallicity from $[Fe/H]<-2.3$ up to solar and above
\citep{perr_cat,morrison},   have opened an important debate and actually gave
rise to the present study. 

In fact, as recently discussed by \citet{morrison}, \citet{beas} and
\citet[][and references therein]{burst04}, the detection of several old
metal-poor clusters with thin-disc kinematics would imply that {\it (i)}  M31
is likely to have had a disc already in place at the very early stages of the
galaxy evolution, and {\it (ii)} no substantial merger event can have occurred
at later epochs as galaxy disc would  have been disrupted or, at least, heated
(but see, however, \citealp{abadi}).  This conclusion is  at odds with the
indications found by \citet{bro03}, who reported the detection of a wide
intermediate-age (6-8 Gyr) population of metal-rich stars (with $[Fe/H]>-0.5$)
in a minor-axis halo field of M31 observed with HST/ACS, interpreted as the
result of the merging with an almost equal-mass companion. {\em As a
consequence, it is of primary interest to verify if and how many clusters do
belong to the claimed thin  disc and,  even more important, how old and
metal-poor they are.}

In the following section we carry out a revision and a new selection of the
candidate sample of BLCCs based on different (and partially complementary)
criteria, relying on the  {\it Bologna Revised Catalog of M31 GCs} (hereafter
BRC; \citealp{silvia}).\footnote{See the latest electronic version of the
catalog available at the Web address:  {\tt http://www.bo.astro.it/M31}} Then,
in Sect.\ 3, we  discuss the properties of the global sample of clusters and in
particular their kinematical properties,  metallicity and estimated ages. 

{\em We anticipate that our analysis lead us to conclude that most (if not
all)  of  the BLCCs are younger than $\sim 2$ Gyr, more metal rich than
$[Fe/H]\sim  -1.0$ and they nicely fit the structural and kinematic
characteristics of  the thin-disc subsystem recently detected by
\citet{morrison}.} This issue is further discussed in Sec.\ 4, where we also
summarize the main results  of our analysis.

\section{Toward a fair sampling of BLCCs in M31}

To carry out a systematic analysis of the BLCC population of the Andromeda
galaxy,  our BRC data have been complemented with the kinematic and spectral
indices  information from \citet{perr_cat}. This choice allowed us to  preserve
full homogeneity in the comparison of line indices, and take advantage at the
same time of a larger sample of clusters and a better accuracy  of radial
velocity measurements. It has to be noted, however, that our main conclusions 
remain essentially unchanged if the \citet{bh90} dataset would have been used
instead.

Throughout the paper, for M31 we will assume a distance modulus  $(m-M)_o =
24.43$~mag \citep{fm90}, a systemic radial velocity $V_r = -301$~km\,s$^{-1}$
and a galaxy center, to which refer the XY coordinate system, both from
\citet{sydbook}.  A single value for the interstellar extinction toward the
galaxy is adopted for simplicity, with $E(B-V)=0.11$ \citep{mcr69,h_book} and a
standard reddening  law (e.g.\ \citealp{scheffler}). We verified that our
results would remain substantially  unaffected when using, in alternative, the
\citet{barmby} reddening values.


\begin{figure*}[!ht]
\centerline{
\psfig{file=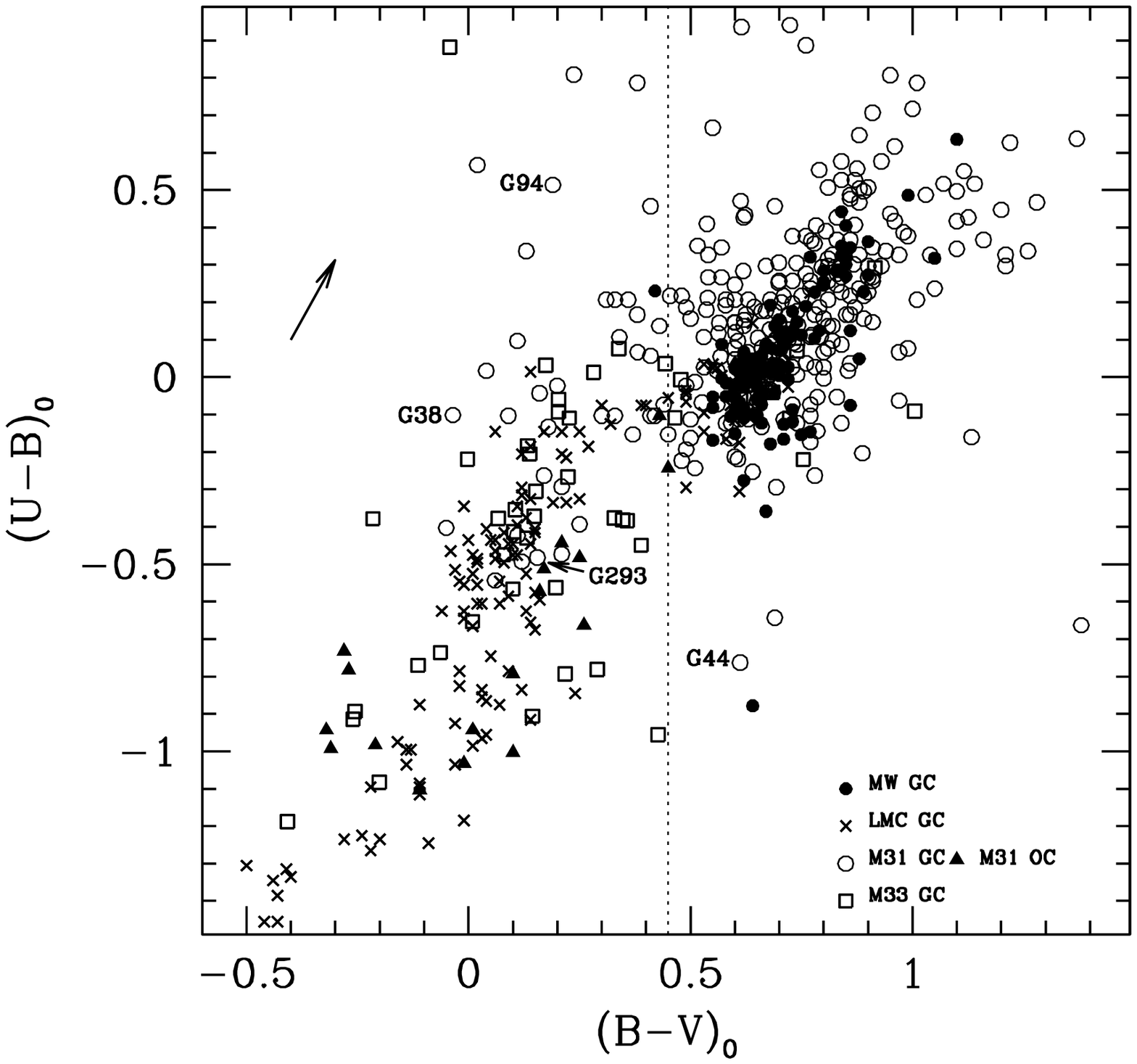,width=0.77\hsize,clip=}}
\caption{Two-color diagram of globular clusters for Local Group galaxies.
Data for M31 globulars are from the BRC \citep{silvia} (open dots), those for the
Milky Way are from \citet{h96} (solid dots), LMC GCs are from \citet{sydMC}(crosses)
and M33 data are from \citet{chandar}(squares). Also reported in the plot are the
M31 open clusters from the \citet{h_open} catalog (solid triangles).
All the data have been reddening-corrected assuming $E(B-V) = 0.11$ for M31, 
0.13 for LMC and 0.07 for M33. MW globulars have been
corrected according to \citet{h96}.
Note the broader color distribution of M31 GCs compared to the MW population.
Vertical line marks the reference value $(B-V)_o = 0.45$, adopted for BLCC selection.
Labeled clusters are those observed by \citet{wh01} with HST. 
The arrow is a reddening vector for $E(B-V)=0.1$~mag.}
\label{bvub}%
\end{figure*}

All the M31 clusters comprised in our analysis (i.e.\ 67 targets collected in
Table~1) belong  to Class~1 BRC entries, that is they are all {\it genuine} 
M31 members confirmed either spectroscopically or by means of high-resolution 
imaging.\footnote{Two original Class 1 BRC objects, namely B430-G025 and NB91,
eventually  resulted foreground field stars to a deeper analysis 
\citep{silvia,beas} and have been excluded from the present sample.} 
The clusters are selected according to their intrinsic color [i.e.\ $(B-V)_o \le 0.45$)]
and/or to the strength of their $H\beta$ spectral index ($H\beta \ge 3.5$~\AA).
The rationale of these selection criteria is explained in detail in the following sections.

\subsection{Color selection}

Figure~\ref{bvub} is a collection of the GC population for  Local Group
galaxies in the reddening-corrected U-B~vs.\ B-V color plane. We joined data
for the Milky Way \citep[][and latest updates]{h96}, the LMC \citep{sydMC}, M33
\citep{chandar}, and M31 \citep{silvia}. For the MW globulars, the adopted
values of the color excess $E(B-V)$ are those originally  provided by
\citep[][]{h96}, while $E(B-V)= 0.13$ for LMC and 0.07 for  M33
\citep{sydbook}.

As a well recognized feature, the large majority of M31 GCs in the figure tend
to bunch  around $(B-V)_o\simeq 0.7$ and closely tracks the locus of the MW
globulars, suggesting a similar age and metallicity distribution.  On the other
hand, as noted long ago by \citet{vet62}, \citet{syd67,syd69}, and  more
recently by \citet{barmby}, several M31 globulars spread over much bluer
colors,  compatible with the young and intermediate-age LMC GCs.\footnote{Note,
however, that some of  the most extreme M31 outliers in Fig.~\ref{bvub} plot
are probably due to poor photometry in one (likely U) of the UBV bands and/or
to inaccurate reddening correction. See \citet{barmby} and \citet{silvia} for a
detailed  discussion of M31 GC photometry.}


\begin{figure*}[!ht]
\centerline{
\psfig{file=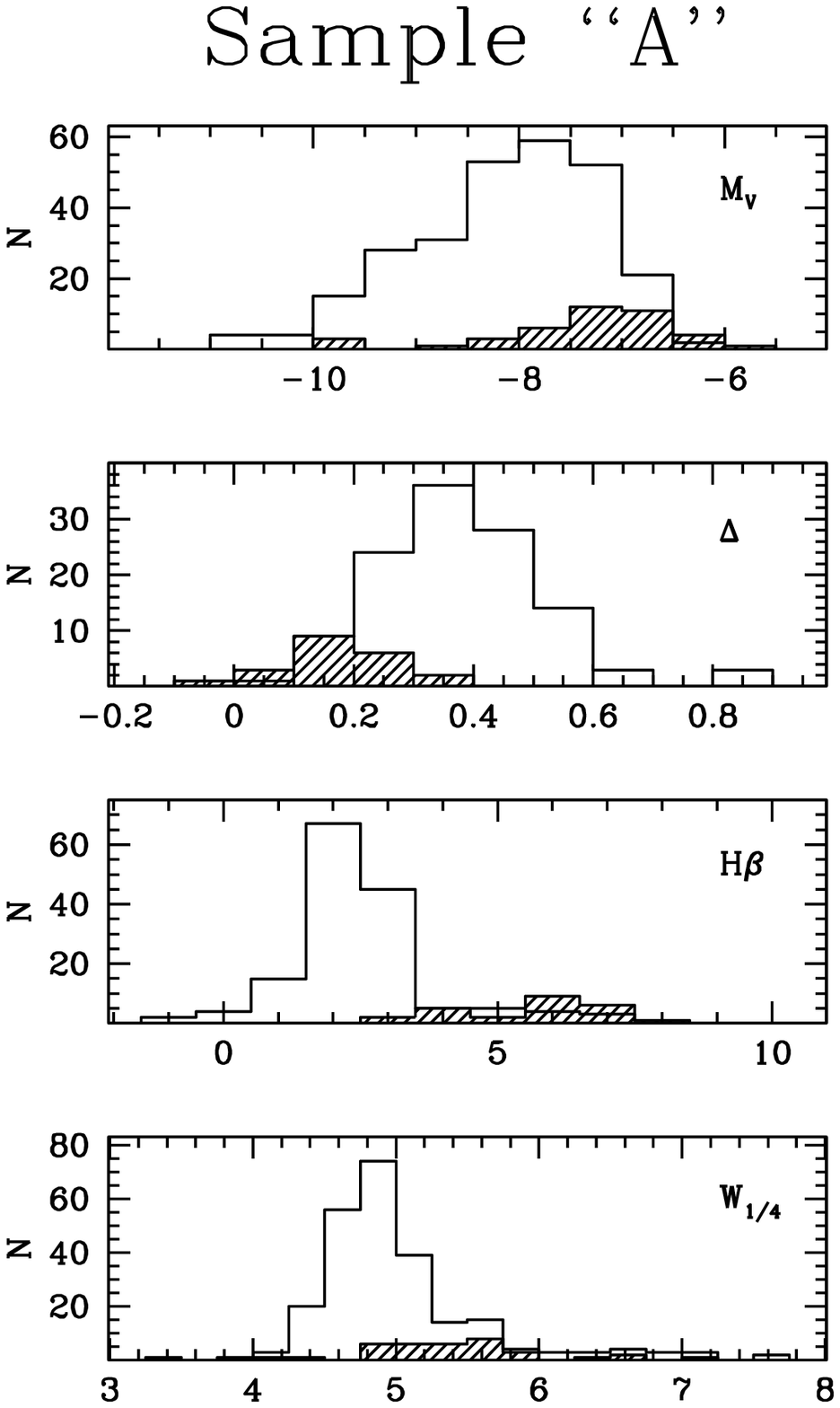,width=0.35\hsize,clip=}
\psfig{file=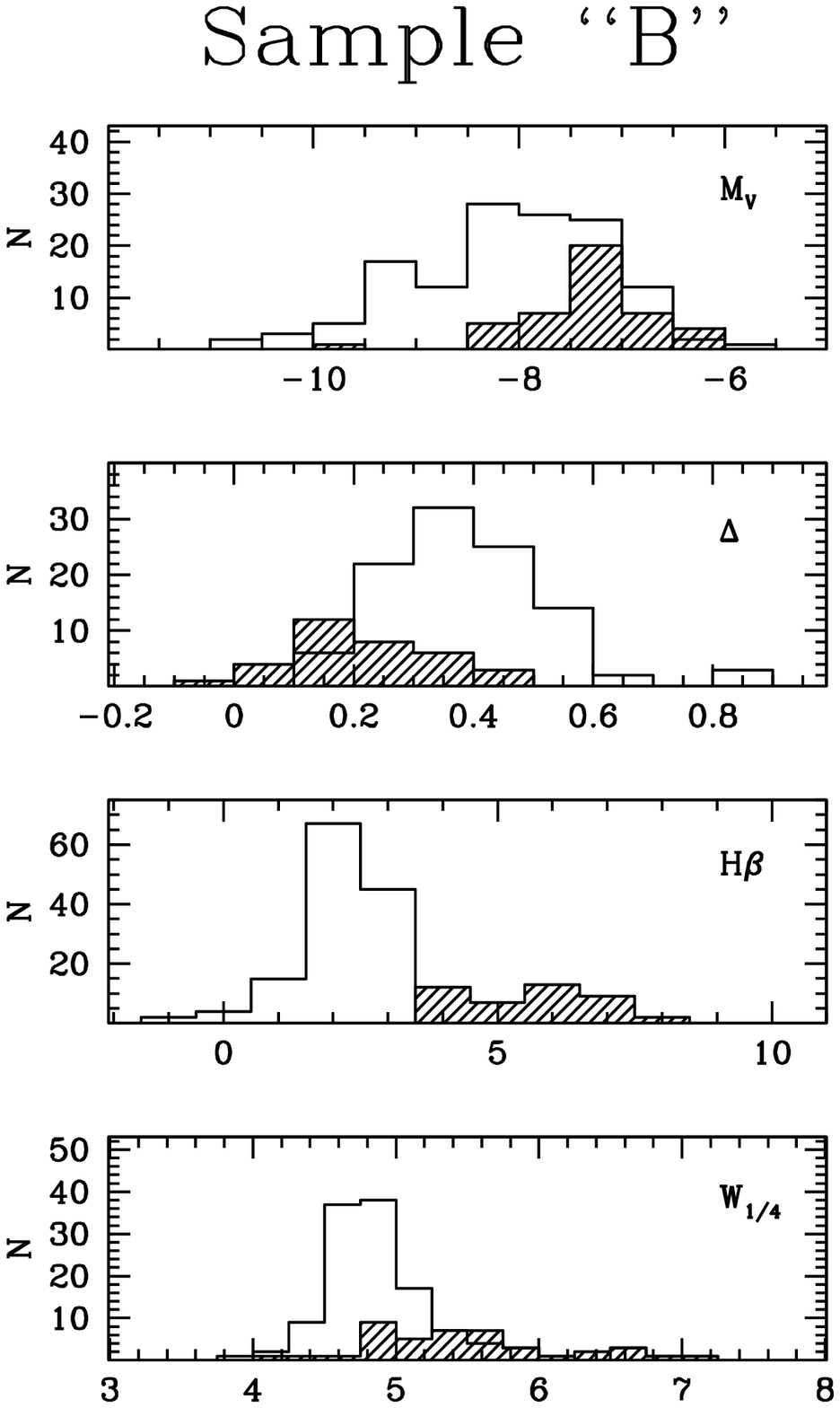,width=0.35\hsize,clip=}} 
\caption{{\it Left panels, from top to bottom} - Distribution of color-selected 
BLCCs (Sample ``A''; shaded histograms) and ``ordinary'' clusters 
with $(B-V)_o > 0.45$ (clean histograms) vs.\  absolute V
magnitude, the 4000~\AA\ Balmer-break index $\Delta$ 
(in magnitude scale, according to \citealp{bh}), 
the Lick H$\beta$ pseudo-equivalent width (in \AA, according to \citealp{faber}), 
and  W$_{1/4}$ morphological parameter (according to \citealp{buon}, and \citealp{bat87}).\protect \\
{\it Right panels, from top to bottom} - Same as the left panels, but for the
H$\beta$-selected BLCCs (Sample ``B''; shaded histograms) vs.\ ``ordinary'' clusters, 
now defined as those with H$\beta < 3.5$  (clean histograms).}
\label{hisbv}%
\end{figure*}


\begin{figure*}[!t]
\centerline{
\psfig{file=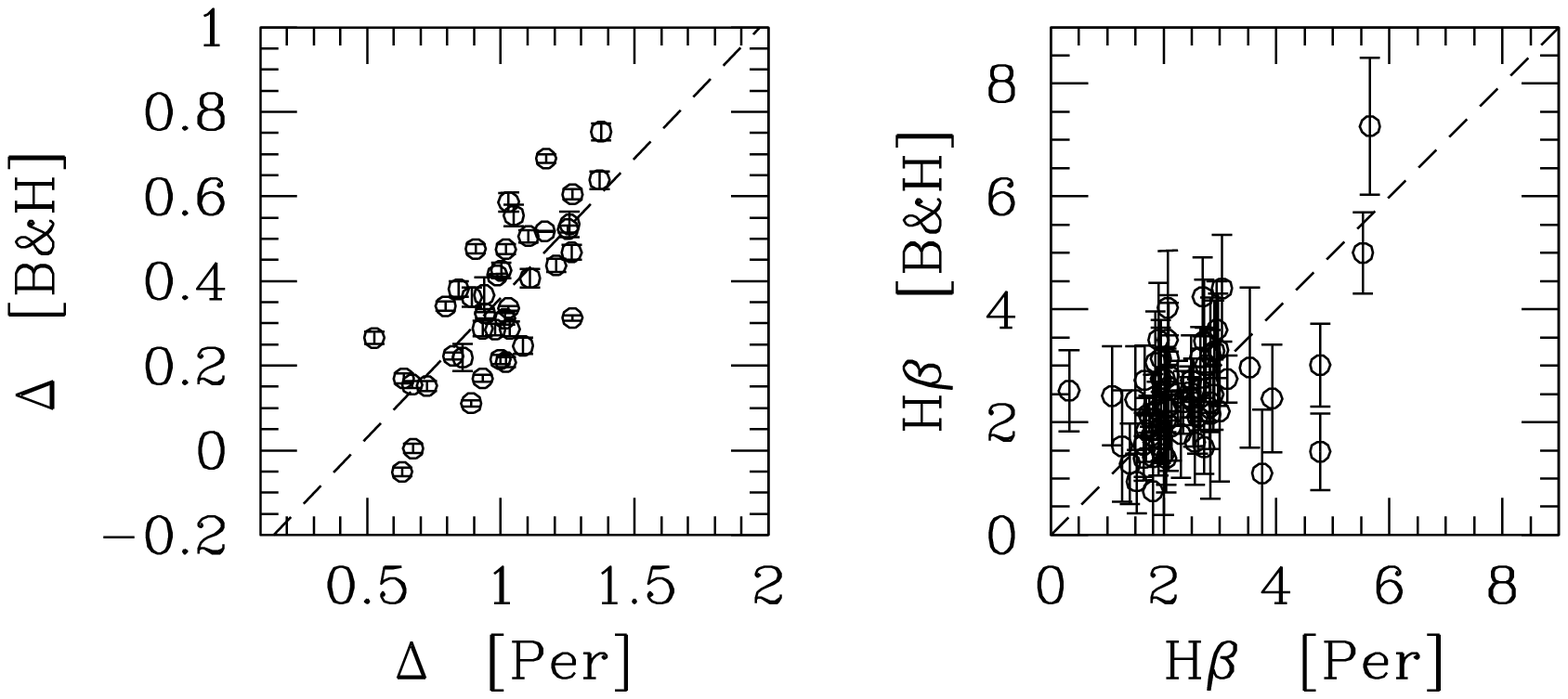,width=0.85\hsize,clip=}}
\caption{Standard calibration for the \citet{bh} $\Delta$ and Lick H$\beta$ indices.
Data for 41 M31 globular clusters of \citet{perr_cat}, in common with \citet{bh90}
are compared, and the least-square equations of transformation are derived, as labeled
in the plots. Note that the \citet{perr_cat} H$\beta$ index has been converted to
pseudo-equivalent width, in \AA, while the 4000~\AA\ Balmer jump index, $\Delta$, is 
expressed in magnitudes.}
\label{calib}%
\end{figure*}

On the basis of the B-V distribution of Fig.~\ref{bvub}, a more immediate  
selection criterion for BLCCs in M31 is to pick up those objects bluer than the
bluest  GC in the MW (i.e.\ NGC~7492, with $(B-V)_o \sim 0.42$; \citealp{h96}).
Operationally, we therefore defined a color-selected BLCC sample (referred to
as Sample ``A'' in our following discussion), consisting of 41 objects (out of
330 confirmed GCs in M31, according to BRC) with $(B-V)_o \le 0.45$. These
objects are reported  in Table~1 (with label ``A'' in column 12); of them, 29
targets have radial velocity estimates in the  list of \citet{perr_cat}.  We
are aware, of course, that the indicative threshold in the integrated $B-V$
color might be a too ``prudent'' selection, as for instance it misses cluster
G44 (see Fig. \ref{bvub}),  claimed by \citet{wh01} to have an age about
100~Myr on the basis of the HST c-m diagram. On the other hand, the other three
young clusters of \citet{wh01} are correctly picked up, and in any case the
relevant uncertainty in the reddening correction for most of the M31 clusters
prevents any firm selection criteron based on  integrated colors alone.
Therefore, Sample ``A'' likely provides a ``clean'' conservative estimate of
the  real fraction of young objects in the M31 GC population, and certainly
secures  our statistics from any contamination of intrisecally ``red''
clusters, even taking into  account the claimed photometric errors, in the
range 0.05 to 0.15 mag, depending on the cluster  location and the source of
the photometry.

\subsection{Blue vs. Red GCs: an overall preliminary comparison }

Besides the obvious difference in color, it is interesting to further
investigate  how Sample ``A'' clusters characterize vs.\ the remaining fraction
of ``red''  M31 GCs. We therefore extended our analysis to different
positional,  morphological and spectrophotometric parameters, according to the
BRC data. Left panels of Fig.~\ref{hisbv} summarize the most striking
differences between  the two cluster populations. The four histograms report
the cluster distribution in absolute $V$ magnitude, in two spectrophotometric
indices, namely the \citet{bh}  $\Delta$ and the Lick  H$\beta$ index
(according to the original definition  of \citealp{faber}), and finally vs.\
the W$_{1/4}$ structural parameter measured  in an homogeneous way by
\citet{buon} and \citet{bat87} for almost all the M31 cluster candidates.

In the plots, absolute V magnitudes from the BRC are the result of a full
revision of  all the available photometry, including the HST data.  Magnitude
differences with respect to the extended database of \citet{barmby} are
typically  less than $\pm 0.2$~mag, depending in general on a different
reddening correction, but with no systematic trend of magnitude  residuals with
GC color. 

Both narrow-band indices considered in Fig.~\ref{hisbv} have been taken from 
\citet{perr_cat};\footnote{\citet{perr_cat} report all indices in magnitude
scale. For the case of H$\beta$, a transformation to pseudo-equivalent width
(EW) units (in \AA) has been done through the equation:  $EW =
27.5\,(1-10^{-0.4\,I})$ \citep[see, e.g.][]{bh90}, where $I$ is the original
index, in magnitude scale.} due to their unfluxed observations, however, the 
instrumental output had to be reduced to a standard system.  This has been done
relying on a set of 41 M31 GCs in common with \citet{bh90}  for which both
$\Delta$ and standard Lick indices are provided in the appropriate  units (see
Fig.~\ref{calib}). As for H$\beta$, which measures the strength of the Balmer
absorption line in pseudo-equivalent width, we verified that no transformation
was necessary for the \citet{perr_cat} original data, that roughly matched the
Lick standard system within a high, but still convenient,  0.9~\AA\ rms. In
case of the $\Delta$ index, measuring the amplitude of the  4000~\AA\ Balmer
jump in the integrated spectra of the clusters, the \citet{bh} standard  system
was reproduced via a linear transformation in the form \begin{equation}
\Delta_{\rm std} = 0.66\,\Delta_{\rm Perrett} - 0.3 \qquad{\rm [mag]}
\end{equation} with a 0.12~mag rms in the individual point scatter. The
corresponding plots in  Fig.~\ref{hisbv} have been corrected to the standard
system accordingly.

Finally, the bottom histogram of Fig.~\ref{hisbv} reports the distribution of
the  \citet{bat80,bat87} W$_{1/4}$ morphological parameter, that nicely relates
to the cluster  core radius through a fitting King profile (see \citealp{bat82}
for further discussion). 

A Kolmogorov-Smirnov test on the data distribution of the left panels of
Fig.~\ref{hisbv}  indicates, at a confidence level better than 99.99\%, that
Sample ``A'' BLCCs differ from the remaining fraction of M31 clusters under all
the considered aspects. Compared to the GC general distribution, BLCCs are
intrinsically fainter, morphologically less concentrated, and with a shallower
Balmer jump. 

The most striking feature, however, is the  intensity of the H$\beta$
absorption line, that is much stronger in BLCCs and  essentially out of the
range covered by ordinary globulars.  This is strongly  suggestive of young
ages, as we will discuss in better detail in the next sections.

\subsection{A different (and complementary) selection: the H$\beta$ index}

The claimed H$\beta$ enhancement for BLCC candidates opens our analysis to a
complementary  selection criterion that, relying on this narrow-band
spectrophotometric index, basically overcomes any problem dealing with the
(poorly known) reddening correction for M31 GCs. The peculiar strength of
H$\beta$ absorption in the spectra of several M31 GCs is a well known feature,
discussed since long ago by \citet{bur}, \citet{trip89}, \citet{bmg},
\citet{lee00}, \citet{per03}, and \citet{schiav},  among others. 
Since homogeneous H$\beta$ values are
available for most of M31 clusters from the \citet{perr_cat} database,  one
could self-consistently rely on this parameter for a more effective diagnostic
of BLCCs.

Similar to the color-selected BLCC sample, on the basis of the observed
distribution of Fig.~\ref{hisbv}, we could try a new sample selection picking
up BLCCs  among those objects with a Lick index H$\beta \ge 3.5$~\AA 
(see Fig.~7 and Fig.~8, below). 
According
to this different criterion, a larger number of clusters (51 in total, with 25
objects in common with the color-selected Sample ``A'') will be included in our
{\it bona fide} BLCC  dataset. This will be our Sample ``B'', as reported in
Table~1 and labelled accordingly in column 12.

Among the 45 objects in Table~1 with complete $B-V$ and H$\beta$ photometry,
only one  Sample ``A'' cluster should not be comprised in the H$\beta$-selected
Sample ``B'' (this is B206-D048, with $(B-V)_o = 0.12$ and H$\beta =
2.53$~\AA), while 19 Sample ``B''  clusters are too red in $B-V$ to be included
in Sample ``A'' (thus confirming the presence of a possibly important fraction
of strongly reddened young clusters in the M31 GC population).

The right panels  of Fig.~\ref{hisbv} complete the comparison between Sample
``A'' and ``B'' properties in the different parameter domain. The substantial
analogy of the two plots and the sharper characterization of the {\it bona
fide} BLCC candidates in the H$\beta$-selected sample makes the latter
spectroscopic selection much more safe and efficient in identifying  the  {\it
same} kind of clusters selected according to the color criterion alone.

\subsection{The final adopted BLCC samples}

From the previous arguments, from now on we will adopt the two samples defined
above and analyse Sample ``A'' and ``B'' objects in parallel. Full information
for these 67 clusters is summarized in Table~1. In particular, photometry and
positional data in the table are from the BRC (unless otherwise  stated),
[Fe/H], radial velocity and spectroscopic indices H$\beta$ and $\Delta$ are
from \citet{perr_cat} (the latter ones, converted according to
Fig.~\ref{calib}), the morphological parameter W$_{1/4}$ is from \citet{buon}
and \citet{bat87}, while $\delta$ kinematic residuals are from
\citet{morrison}.

As we already stressed before, a color selection like for Sample ``A'' better
favours a plain observational approach (90\% of the objects in Table~1 has a
measured $B-V$, while for only 78\% H$\beta$ is provided from spectroscopy),
although it is prone to any uncertainty in the reddening correction; on the
other hand, H$\beta$ selection, such as in Sample ``B'', likely provides a more
confident and physical BLCC selection, but suffers from more difficult 
observing constraints. {\it Quite importantly, however, we will show that,
disregarding any preferred selection  criterion, the same conclusions hold for
both samples.}

As a final remark to our sample selection, we should also note that a number
of  other BLCC candidates (confirmed or not) have been picked up by various
authors based on different observations and procedures. For the sake of
completeness, we have carried out an exhaustive search in the M31 literature
assembling a coarser set of reportedly ``young'' objects, listing our results 
in Table~2 (21 confirmed GCs) and Table~3 (69 GC candidates to be confirmed),
together with a note to the corresponding reference studies.\footnote{Data
sources for Table~2 and 3 entries are the same as for Table~1 unless
explicitely reported. Note, of course, that for Table~3 targets  one would
require a clear-cut spectroscopic check due to a possibly high contamination by
spurious objects.}

In the present analysis, however, we will restrain only to the {\it bona fide}
samples  (i.e.\ Table~1 data), postponing to a future paper a complete review
of each individual object.


\begin{figure*}[!t]
\centerline{
\psfig{file=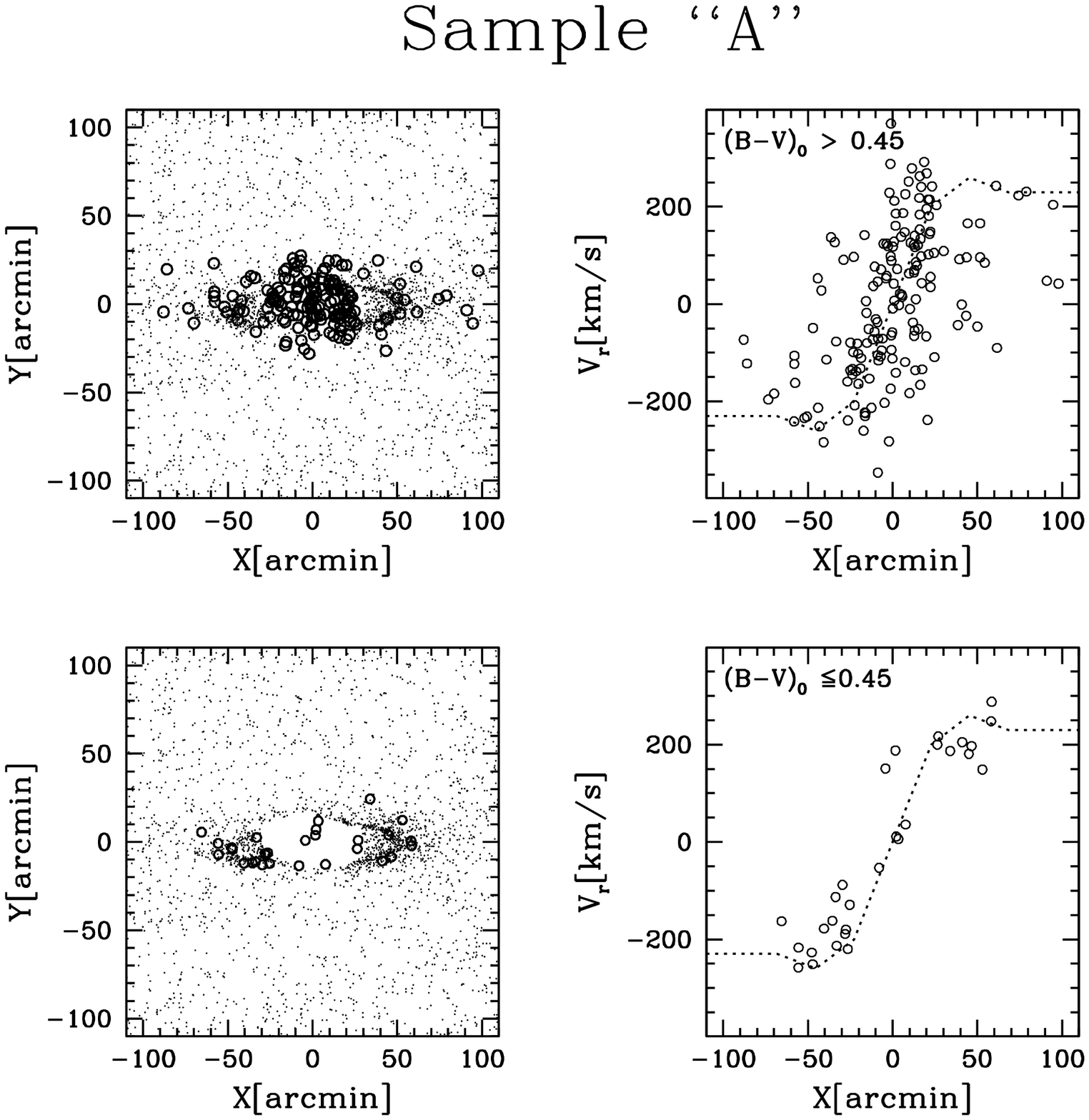,width=0.5\hsize,clip=}
\psfig{file=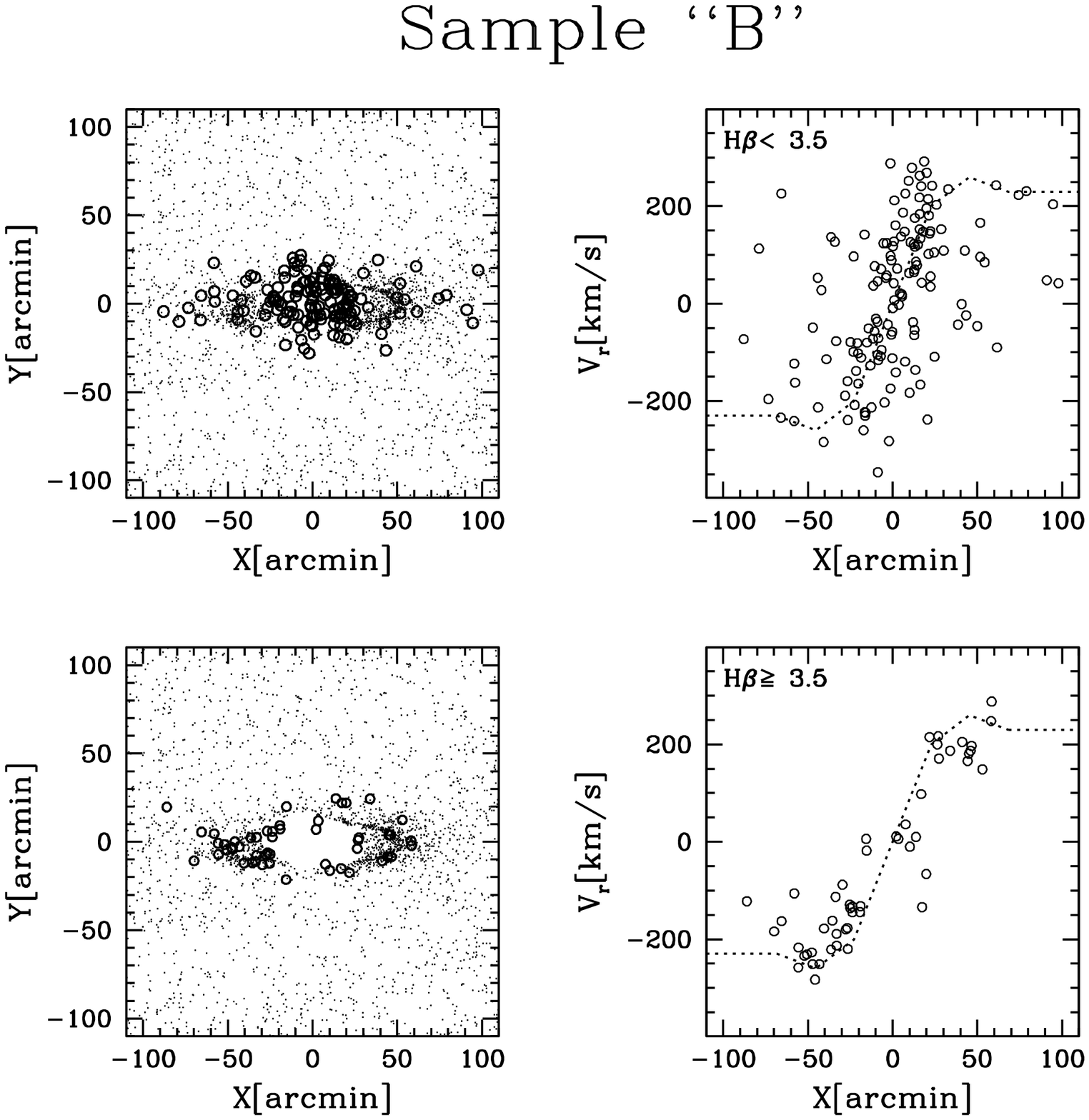,width=0.5\hsize,clip=}}   
\caption{Spatial distribution and kinematic properties of M31 ``ordinary'' globular
clusters, with $(B-V)_o > 0.45$ or H$\beta < 3.5$~\AA, as labeled {\it (upper panels)}, compared
with the BLCCs of Sample ``A'' and ``B'' {\it (lower panels)}.
The smoothed disc structure in the M31 maps is traced by the PSC-2MASS sources \citep{cutri},
mainly consisting of bright AGB members and Carbon stars. Overplotted on the radial velocity 
distribution, as a function of the major axis displacement X, is the rotation curve of M31
disc according to \citet{sydbook}.}
\label{kine}%
\end{figure*}

\section{Dynamics and distinctive physical properties of BLCCs}

After a thorough analysis of the positional and kinematical properties of M31
GCs, based on the \citet{perr_cat} data, \citet{morrison}  concluded that the
GCs in this galaxy belong to two kinematical components: ``...a thin,  rapidly
rotating disc, and a higher velocity dispersion component whose properties
resemble that of the Bulge of M31.'' According to these authors, membership to
one of the two dynamical components is assessed for each cluster on the basis
of its residual velocity with respect  to a disc model, picking up disc
clusters among those with residual $|\delta|<0.75$ (in normalized units, see
\citealp{morrison}).  The claimed probability of mis-classification with this
procedure turns to be about 30\%.


\begin{figure*}[!t]
\centerline{
\psfig{file=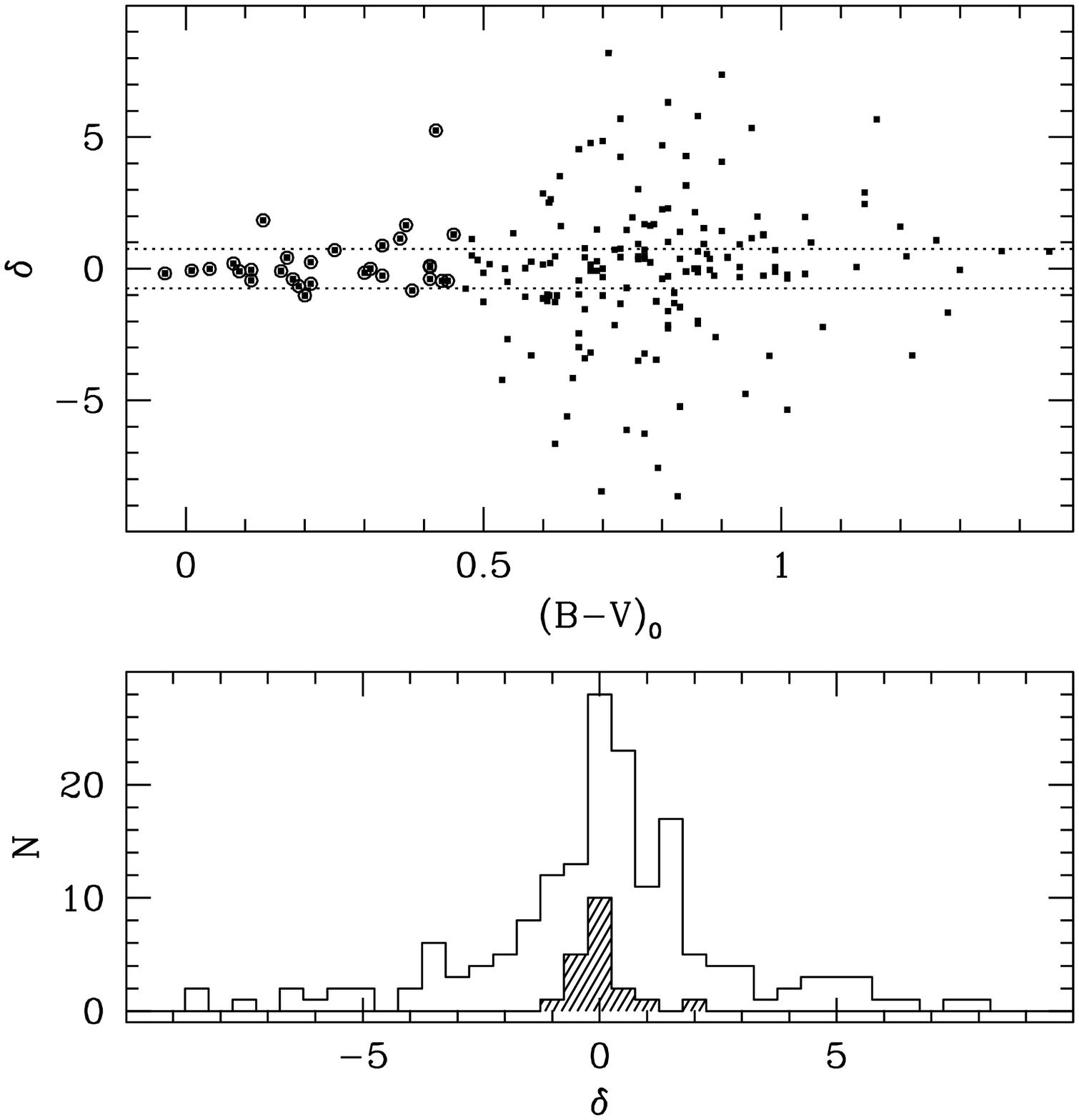,width=0.4\hsize,clip=}
\psfig{file=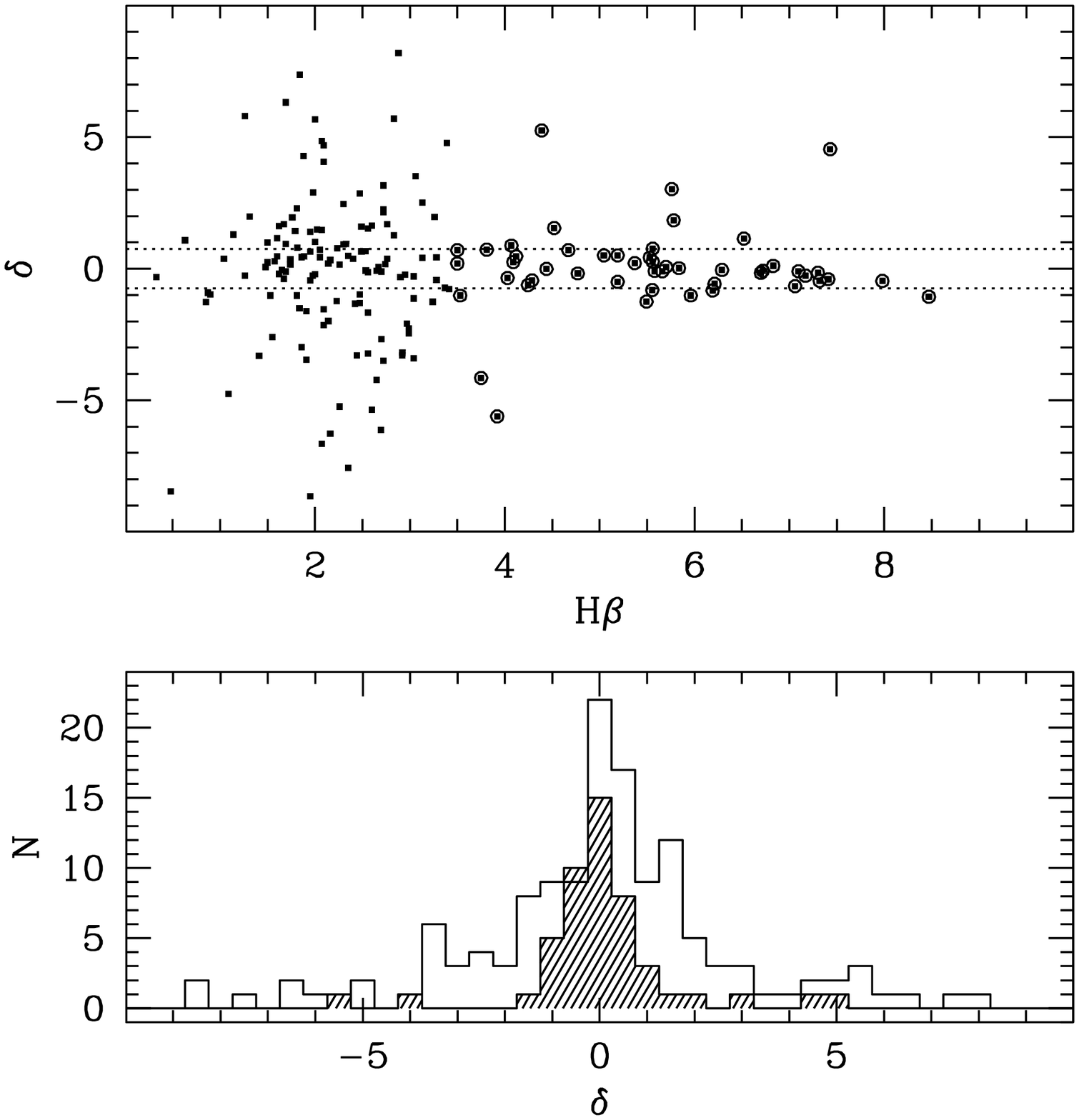,width=0.4\hsize,clip=}}
\caption{GC kinematic residuals, $\delta$, with respect to
the \citet{morrison} disc model for M31. The BLCCs distribution, according
to Sample ``A'' {\it (upper left panel)} and ``B'' {\it (upper right panel)} is marked on the plots (big solid dots)
compared to the remaining fraction of ``ordinary'' clusters (small points).
The dotted strip encloses the low-residual region with $|\delta| \le 0.75$ (such
low-$\delta$ clusters basically share the disc kinematic, according to \citeauthor{morrison}'s model).
The histogram of the $\delta$ distribution for each plot is also summarized in 
the lower panels for BLCCs (shaded) and ``ordinary'' (clean histograms) GCs.}
\label{residuals}%
\end{figure*}

\subsection{The Morrison et al.\ (2004) framework and  the spatial and
kinematic properties of BLCCs}

In Fig.~\ref{kine} we compare the spatial distribution (in the X and Y
projected distances along the major and minor axes, respectively) and the $V_r$
vs.\ X distribution of Sample ``A'' (left panels) and ``B'' (right panels) 
BLCC candidates with the remaining fraction of ``ordinary'' M31 globulars. The
thin plot of points in the X vs.\ Y panels mainly maps the brightest AGB  and
Carbon stars of M31, from the 2MASS Point Sources Catalogue
\citep{cutri},\footnote{The 2MASS sources with color and magnitude constraints
such as  $1.3 \le J-K_S \le 2.0$ and $14.0 \le K_S \le 15.7$ have been selected
to better tune on  M31 stellar population.}  and gives therefore a neat picture
of the outermost structure of galaxy disc (\citealp{sydbook,h_book}), allowing
us to appreciate in some detail  the overplotted GC distribution.

It is evident from the figure that ordinary clusters (upper panels) are
uniformly  distributed  all over the apparent body of M31, up to distances of
$X \simeq 100\arcmin$   ($d \simeq 22$~kpc) from the galaxy center, closely
tracing the smooth luminosity  profile of M31. In addition, most of the bulge
clusters show clear  sign of coherent rotation in the $V_r$ vs.\ $X$ plane, as
a part of the rotationally supported structure of the galaxy \citep{morrison}. 
However, the large scatter around the overplotted rotation curve \cite[taken
from][see his Fig.~3.10]{sydbook}, indicates that pressure support has a 
significant role in the overall kinematics of the sample (see, however,
\citealp{morrison} for a deeper analysis of these clusters).

On the contrary, BLCCs  seem to avoid the inner regions of the galaxy  (see
lower panels in Fig.~\ref{kine}) and appear well projected onto the outer disc
with  a strong correlation with the underlying spiral substructures. Also the
velocity pattern quite well traces the velocity curve of the disc. 

{\it Our conclusion, therefore, is that both color-selected and
H$\beta$-selected BLCCs belong to the cold thin disc of M31, in agreement with
previous  results of \citet{beas} and \citet{burst04}, that were limited, 
however, to a much smaller subset of clusters.}

To further clarify this issue, in the upper panels of Fig.~\ref{residuals} we
plotted the \citet{morrison} residuals (taken from Table 2 therein), singling
out our {\it bona fide} BLCCs. It is quite evident from the plots that BLCCs
stand out naturally as a very low-$\delta$ family.  Almost the whole BLCC
population (independent of the adopted selection), in fact, matches  the
\citet{morrison} disc-membership criterion, and only a few such clusters
exceed  $|\delta| > 1.5$. This feature is summarized also in the lower panels
of Fig. \ref{residuals}, where the shaded histograms give the residual
distributions of BLCCs compared to the whole sample of ``ordinary'' GCs.
Complementing the \citet{morrison} results, we can conclude that

\noindent$\bullet$  {\it both color- and  H$\beta$-selected BLCCs can
effectively comprise a significant  fraction of ``disc clusters'', being
natural tracers of the cold thin disc  subsystem of M31 (\citealp{morrison},
but see also \citealp{burst04}).} 

\noindent$\bullet$  {\it BLCCs appear to constitute a separate family for what
concerns  a number of non-kinematical properties like, for instance, the
absolute magnitudes, the H$\beta$ and $\Delta$ indices (at least), and the
structural parameters.}


\begin{figure}[!t]
\centerline{
\psfig{file=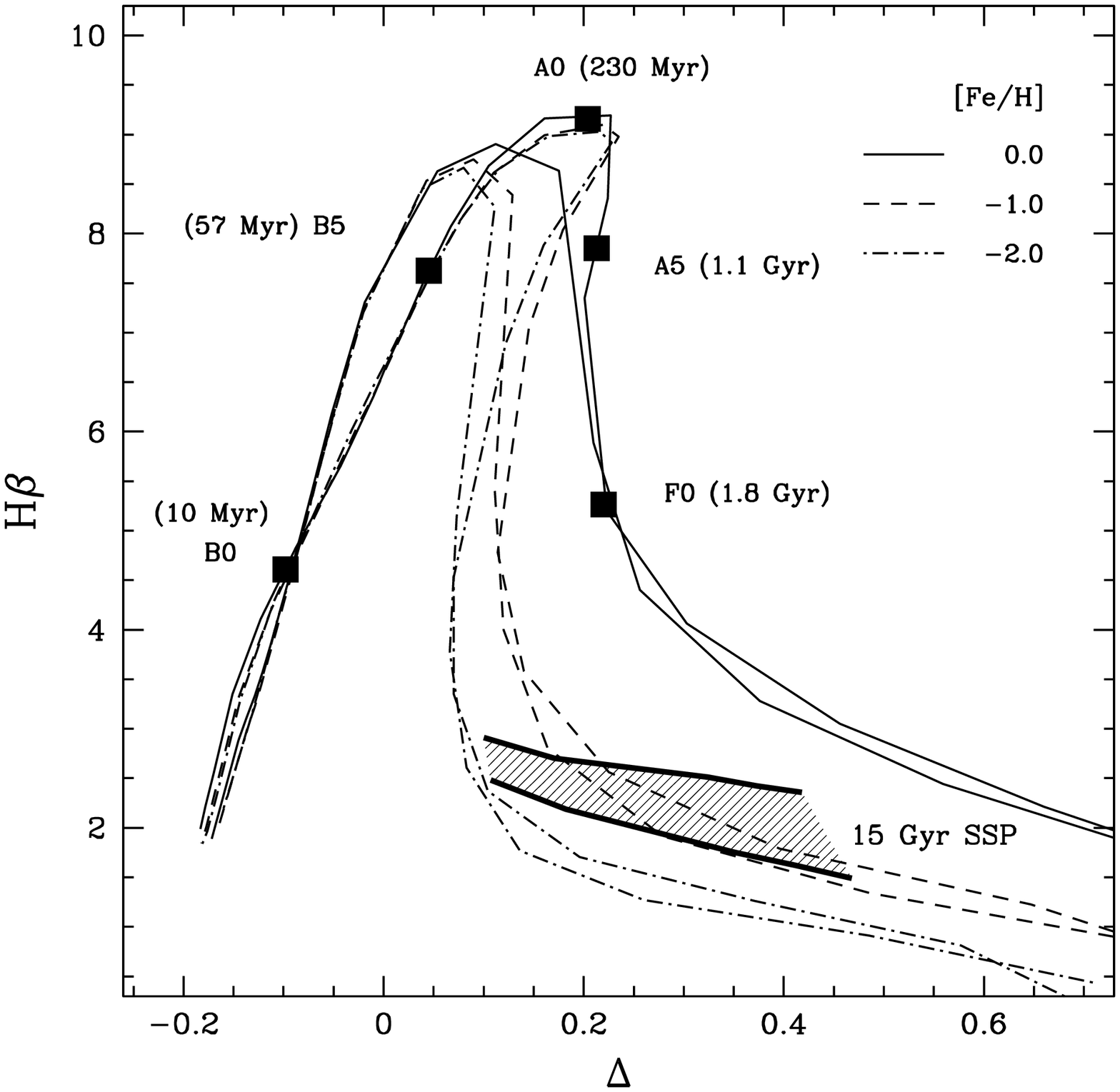,width=\hsize,clip=}}
\caption{Theoretical diagram for the H$\beta$ Lick index (in \AA) vs.\ Balmer jump $\Delta$ index
(in magnitude scale).
Thin curves are the expected locus for MK~V stars of different temperature, from 5000 to 50\,000~K,
surface gravity ($\log g = 4$ and 5, the latter being the shelf of curves peaking at slightly 
higher values of H$\beta$), and metallicity (from $[Fe/H] = -2$ to solar, as labeled).
Theoretical indices rely on the {\sc Bluered} library of synthetic stellar spectra \citep{bert1,bert2}. 
The empirical spectral-type vs.\ effective temperature calibration  of \citet{j66}
is marked on the $(\log g, [Fe/H]) = (5, 0.0)$
curve, together with the estimated main sequence lifetime of stars, as derived from
\citet{b02}. The shaded strip is the region for the 15 Gyr SSP models of \citet{buz89},
with different  HB morphology. Lower edge in the strip is for a red-clump HB morphology, while upper edge 
is for a blue HB. SSP metallicity spans the range $[Fe/H] = -2.27 \to +0.22$~dex, in the 
sense of increasing $\Delta$.}
\label{hbd_teo}%
\end{figure}

\subsection{The H$\beta$ and $\Delta$ indices as age tracers}

In a quite common view, a blue integrated color for a cluster is actually an
indication of a young age as its integrated color is dominated by the bright
blue stars at the MSTO point. On this basis, even the early studies of the M31
clusters (see e.g.\ \citealp{syd67,syd69}) immediately noticed that most of the
blue cluster candidates should be young. The combined information provided by
H$\beta$ and the Balmer jump, $\Delta$, allow in principle to tackle the
problem of dating BLCCs on finer detail; both indices, in fact, are contributed
by the warmer stellar component in a  simple stellar population (SSP), and can
selectively probe the MSTO temperature  location \citep{b95a,b95b}, leading
therefore to an indirect estimate of age in a stellar aggregate.

Note that, for its photometric characteristics, the $\Delta$ index recalls, to
some extent, the broad-band U-B color, but its narrower wavelength baseline
makes the \citet{bh} index much less sensitive  to the reddening and therefore
a better tracer of the intrinsic properties of  a stellar cluster.\footnote{The
\citet{bh} $\Delta$ index is defined in a magnitude scale as $\Delta =
2.5\,\log (F2/F1)$, where $F1$ and $F2$ are  the luminosity densities (per unit
wavelength) in two 200~\AA\ wide bands centered at 3900 and 4100~\AA,
respectively.  This has to be compared with the Johnson U and B bands, centered
respectively at 3600 and 4400~\AA. According to  \citet*{scheffler}
compilation, the expected color excess  can be computed as $E(\Delta)/E(U-B) =
0.25$ and $E(\Delta)/E(B-V) = 0.18$. \label{foot}} 

Figure~\ref{hbd_teo} will be our reference plot to summarize the main features
of  a H$\beta$ vs.~$\Delta$ diagram, from the theoretical point of view.  In
the figure we reported the expected locus  for main sequence stars of different
temperature, up to $T_{\rm eff} = 50\,000$~K and their inferred spectral type,
according to the \citet{j66} calibration, as labeled.

Our calculations rely on the new {\sc Bluered} theoretical library of stellar
SEDs  \citep{bert1,bert2}, from which we extracted the grid of model
atmospheres with  $\log~g = 3$ and 5, and [Fe/H] = --2.0, --1.0 and solar. The
vanishing sensitivity of both spectral indices to the contribution of late-type
stars makes the H$\beta$ vs.~$\Delta$  information nearly independent of the
exact post-MS details of the c-m diagram,  as far as we compare with the GC
data. This eases a very immediate and straightforward comparison  of the
theoretical stellar locus with the integrated indices of BLCCs, providing a
first  effective constraint to age, through a MSTO calibration linking  $T_{\rm
eff} \to M_{MSTO} \to t_{\rm MS}$ \citep[see, e.g.][]{b02}.

In a more elaborated scheme relying on SSP models, however, one should also
account for  the supplementary contribution of blue HB stars, that are expected
to play a role at some stage of late evolution of low-mass stars modulating the
integrated SED of old stellar populations, typically beyond 10~Gyr. A more
accurate prediction of the integrated output for SSPs of different metallicity
and HB morphology is reported in the figure, according to the \citet{buz89} 
population synthesis models.

The effect of HB morphology is sketched for the 15 Gyr SSP models mainly
resulting in a sensible enhancement of the H$\beta$ strength. It is clear,
however, that in no cases the late evolutionary scenario could account for very
strong H$\beta$-lined clusters (i.e.\ H$\beta \gtrsim 3.0$~\AA),  where a
prevailing contribution of F-type stars calls for a genuinely blue MSTO, and
therefore  for a young age, certainly less than a few Gyrs.


\begin{figure}[!t]
\centerline{
\psfig{file=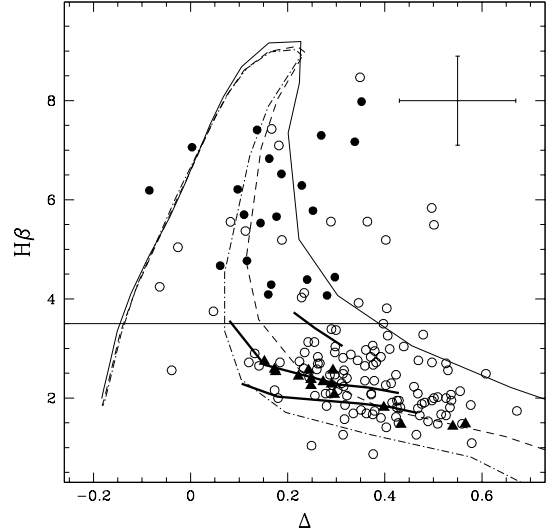,width=\hsize,clip=}}
\caption{The M31 GC distribution in the H$\beta$ vs.\ $\Delta$ index plane. Reference curves
for $\log g = 5$ stars of different temperature (from 5000 to 50\,000~K) and metallicity
([Fe/H] from --2 to solar) are reported from Fig.~\ref{hbd_teo}. Sample ``A'' (i.e.\ color-selected)
BLCCs are singled out (solid dots). By definition, Sample ``B'' comprises all the targets in the plot 
above the H$\beta = 3.5$~\AA\ threshold, as marked (40 objects with available indices 
from Table~1). For comparison, solid triangles are
the \citet{bh90} data for MW GCs, while thick solid lines are the locus for the \citet{buz89} SSP 
models with $t = 15$, 8 and 2~Gyr (in the sense of increasing H$\beta$), red HB morphology, and 
metallicity $[Fe/H] = -2.27 \to +0.22$. Typical error bars for M31 data are reported top left.}
\label{hbd}%
\end{figure}

The distribution of M31 GCs in the H$\beta$ vs.\ $\Delta$ plane is displayed
in  Fig.~\ref{hbd}. Big solid dots are the 21 Sample ``A'' BLCCs from Table~1
with available  H$\beta$ and $\Delta$ indices (by definition, Sample ``B''
comprises all the 40 solid and  open dots above the H$\beta = 3.5$~\AA\
threshold, as indicated in the figure). The MW GC data from \citet{bh90} (solid
triangles) are also reported for comparison.  The stellar loci of
Fig.~\ref{hbd_teo} (only those for $\log g = 5$, and different metallicity, for
the sake of clarity) have also been overplotted, together with the SSP locus 
for different age (and a red HB morphology), according to \citet{buz89}.

It is evident from the figure that, while the bulk of ordinary GCs (including
the MW globulars) has spectral properties consistent with an age of several
Gyrs, the BLCCs indices are clearly dominated by a younger stellar component of
A-F stars of moderately high metallicity  (mostly solar or slightly sub-solar).
This places a confident upper limit to  BLCC age at $\sim 2$~Gyr.
Note also (Fig.~8) that only 4 on 40 BLCCs lie on the branch of the
theoretical stellar loci typical of ages lower than $\sim 50$ Myr
($\Delta<0.0$), while the bulk of the population is consistent with ages larger
than $\sim 200$ Myr.


\begin{figure}[!t]
\centerline{
\psfig{file=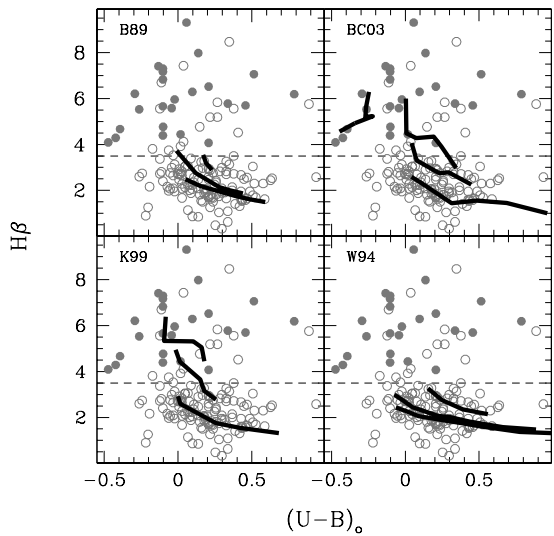,width=\hsize,clip=}}
\caption{The M31 GC distribution in the H$\beta$ vs.\ (dereddened) $U-B$ index plane. 
Like in Fig.~\ref{hbd}, Sample ``A'' BLCCs are singled out by solid dots while, by definition, 
Sample ``B'' comprises all the objects with H$\beta \ge 3.5$~\AA.
The four panels report a comparison with SSP models from different population synthesis codes,
according to \citet{buz89} ({\it B89, upper left panel}, for $t = 15$, 8 and 2~Gyr), \citet{bruz} 
({\it BC03, upper right}, for $t = 15$, 2, 1 and 0.1~Gyr), \citet{k99} ({\it K99, lower left}, 
for $t = 15$, 2 and 1~Gyr), and \citet{w94} ({\it W94, lower right panel}, for $t = 12$, 8 and 2~Gyr).
In all cases, synthetic H$\beta$ index increases with age (excepting the 100~Myr model
of \citealp{bruz}, with a negative $U-B$ and a ``fading'' H$\beta$ index, recalling the 
trend of Fig.~\ref{hbd_teo}.}
\label{hbub}%
\end{figure}

\subsubsection{The $(U-B)$ color as an alternative to the $\Delta$ index}

As we have been commenting in previous section, a more standard assessment of
the BLCC  distribution with respect to the overall GC population in M31 can be
carried out relying  on the $U-B$ color. For its nature, this choice would
allow a more comfortable match with the theoretical output of population
synthesis models, and certainly ease the inclusion of  a larger GC database
with available data, although observations are obviously more critically
plagued by reddening uncertainty (cf.\ footnote~\ref{foot}).

In Fig.~\ref{hbub} the M31 (dereddened) data are compared with the expected SSP
evolution  according to four different population synthesis codes, namely
\citet[and further Web updates]{buz89}, \citet{w94},  \citet{k99}, and
\citet{bruz}. Again, models agree in the overall classification scheme, with
the age of H$\beta$-poor GCs fully consistent with old (i.e. $10-15$~Gyr) 
evolutionary scenarios and BLCCs better matched by young ($t \lesssim 2$~Gyr) 
SSPs.\footnote{In this framework, the peculiar   case of B206-D048, i.e.\ the
only Sample ``A'' cluster not comprised in the BLCC H$\beta$  selection (see
Sec.\ 2.3), might be noticing an old and very metal-poor composing  stellar
population. This interesting cluster does not appear in Fig.~\ref{hbd} and
\ref{hbub}  as it lacks $U-B$ and $\Delta$ indices (see Table~1).}

The emerging evidence, from the combined analysis of the BLCC age range
(Fig.~\ref{hbd} and  \ref{hbub}), coupled with the kinematical information of
Fig.~\ref{kine} leads eventually to the important conclusion  that {\it over
25\% of the \citet{perr_cat} sample (that is at least a 15\% of the whole
sample of confirmed  M31 clusters) consists  of young BLCCs, located in the
outskirts of the thin  disc of M31 and sharing the kinematic properties of this
galactic component.} This global evidence is quite new and points to the
existence of a significant  difference between the M31 and MW globular cluster
systems, which may have deep implications on the understanding of formation and
evolution of the two parent galaxies.

\subsection{Metallicity: BLCCs might be not so metal-poor as claimed}

Another important issue in the \citet{morrison} discussion concerns the claimed
similarity of ``disc'' and ``bulge'' GC metallicity that, like in the MW,
should be regarded as a striking sign of coevality for the whole GC population
in M31. In fact, if our {\it bona fide} BLCCs were mostly metal-poor, as
apparently deduced from the metallicity values reported by \citet{perr_cat},
one would be left with the embarrassing  scenario where the vast majority of
the most metal-poor clusters in M31  should be young and likely members of the
thin-disc galaxy subsystem. Though not impossible in principle, this conclusion
deserves however a deeper check. 

\begin{figure}[!t]
\centerline{
\psfig{file=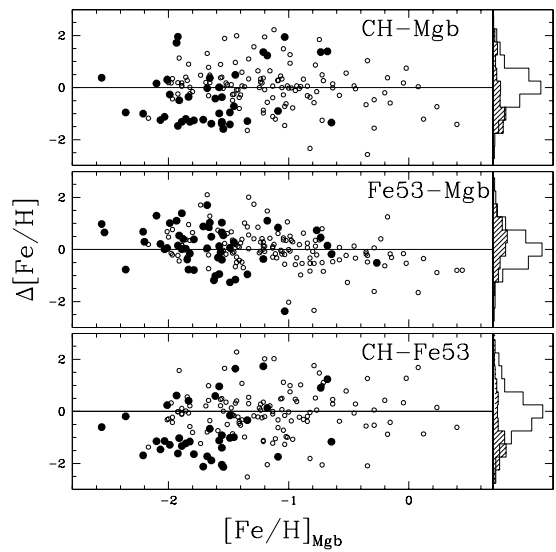,width=\hsize,clip=}}
\caption{Metallicity estimates from single Lick spectrophotometric indices for M31 GCs
according to \citet{perr_cat}. Reported are the differences in the inferred value of
[Fe/H] for each individual cluster from the empirical calibration of the CH (G band), Mgb 
and Fe5335 indices. Sample ``B'' BLCCs are singled out by big solid dots, while
$\Delta [Fe/H]$ residual distribution is summarized by the vertical histograms on the right
(shaded curve for BLCC distribution and clean histogram for the remaining GC population).
For BLCCs it is evident a systematic bias on the average estimate of [Fe/H]
(reported in Table~1) induced by the exceedingly metal-poor abundance 
inferred by the G band ($\Delta [Fe/H] \simeq -1$~dex, cf.\ upper and lower panels 
in the figure). Both Mgb and Fe5335 metallicity values are on the contrary fully self-consistent, 
in average (cf.\ middle panel).}
\label{dmetal}%
\end{figure}

The \citet{morrison} metallicity scale comes from the \citet{perr_cat} [Fe/H]
estimates,  relying on the empirical calibration of three narrow-band Lick
indices (namely, the G-band,  Mgb and Fe5335) vs.\ a coarse set of M31 GCs with
available [Fe/H] determination from  the literature. Given the prevailing
contribution of {\it old} GC in this sample,  the \citet{perr_cat} calibration
actually introduces a subtly circular bias in the  inferred value of [Fe/H] for
BLCCs due to  the well recognized age-metallicity  degeneracy \citep{rb86},
that makes young metal-rich clusters to resemble old metal-poor ones.  This is
shown in the three panels of Fig.~\ref{dmetal}, where we  plotted the [Fe/H]
difference as inferred from the Lick index determinations for each individual
cluster.


From the figure one can immediately recognize a systematic bias induced by the
[Fe/H] estimates from the G-band. As a matter of fact, while the metallicity
scale, as derived from the Fe5335 and Mgb indices, basically agrees in average
(see middle panel in Fig.~\ref{dmetal}),  for the BLCCs (solid dots in the
plots) the G-band line strength tends to yield a significantly lower (roughly $\sim
1$~dex) value of [Fe/H].  This systematic trend especially affects BLCCs, whose 
low age mimics the effect of metal deficiency.

In a SSP, this CH-molecular feature maximizes its sensitivity  to the coolest
stars at the tip of the Asymptotic (AGB) and Red giant branches (RGB)
\citep{gorgas,bur}, and is therefore a mixed age/metallicity indicator, 
through a composite and complex dependence on the AGB morphology. Conversely,
both the  Mg and Fe indices are better sensitive to the bulk of stars in the
RGB, via chemical  opacity, and more confidently trace SSP metallicity
\citep{b95a,b95b}. 

As a further argument, we note that even the observed morphology of the
\citet{wh01} c-m diagrams for the four BLCCs observed so far with HST seem to
indicate  a metallicity in the range $-0.7\lesssim [Fe/H]\lesssim 0.0$.  This
also basically agrees with our ``first-look'' estimate of [Fe/H] from
Fig.~\ref{hbd}, where the BLCCs H$\beta$~vs.~$\Delta$ distribution suggests  a
moderately enhanced metallicity (i.e.\ $[Fe/H] \gtrsim -1.0$~dex).


\begin{figure}[!t]
\centerline{
\psfig{file=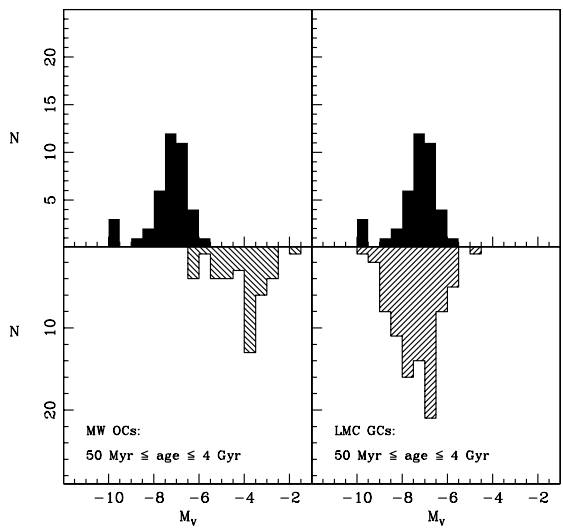,width=\hsize,clip=}}
\caption{Comparison of the Luminosity Function (LF) of M31 BLCCs (filled
histogram, upper panels) with: 
(left panel) the LF of the Open Clusters (OC) of the Milky Way in the same age
range;
(right panel) the LF of LMC globular clusters in the same age
range.
}
\label{oc}%
\end{figure}

\subsection{BLCCs: blue globulars or massive open clusters?}

The actual classification of BLCCs as "young globular clusters" or "massive open
clusters" is somehow a question of semantics. The real point that is worth to
investigate is if these clusters have a counterpart in our own Galaxy, e.g. if
clusters of similar age {\em and} luminosity do exist in the Milky Way.
While the luminosity range spanned by BLCCs is comprised within that of
ordinary globular clusters,
the age distribution of present-day MW globulars is obviously not consistent
with the young age of BLCCs. On the other hand, Galactic Open Clusters (OC) are
comparably young but they appear less luminous on average than BLCCs. 

The Luminosity Functions (LFs) of the Sample B clusters and of the Galactic OCs
\citep[data drawn from the WEBDA database]
[see {\tt http://obswww.unige.ch/webda}]{webda} are compared in the left 
panel of Fig.~\ref{oc}. The plot shows that Galactic OCs with ages similar
to BLCCs are sistematically fainter, the two histograms barely overlaps.
The only Galactic OCs that reach the luminosity range covered by BLCCs are
younger than 30 Myr (e.g. they are clusters whose luminosity budget is dominated
by a few massive stars, much different f.i. from the BLCCs studied by \citet{wh01}), 
while Fig.~7 and Fig.~8 above indicates that $\simeq 90$ per cent
of BLCCs is likely older than $\sim 200$ Gyr. Hence, regardless of the
quite high incompleteness that probably still affects the LF of M31 clusters for
$M_V\ge -6.0$, the Milky Way lacks OCs as luminous as BLCCs in the proper age 
range  (200 Myr $\le$ age $\le$ 2 Gyr). 

Conversely, the right panel of Fig.~\ref{oc} show that the luminosity range spanned 
by M31 BLCCs is very similar to that covered by LMC globular clusters of similar 
age \citep[data from][]{sydMC}. 
The above direct evidences lead to conclude that there is no Galactic counterpart to the young
massive M31 clusters studied here; they are much younger than present-day
Galactic globulars and they are much more luminous than 
present-day Galactic OCs of similar age. On the other hand BLCC counterparts
are quite common in the LMC. Obviously, this conclusion rests on the age estimates of BLCCs
as derived from Fig.~\ref{hbd}, above. The involved uncertaintes leave (formally)
open the possibility that several BLCCs have ages $< 50$~Myr. If so, they should be
interpreted as the counterparts of young open clusters of the Milky Way. The availability of 
a deep c-m diagram is probably the only observational test that can eventually establish the
real nature of these objects.

\begin{figure}[!t]
\centerline{
\psfig{file=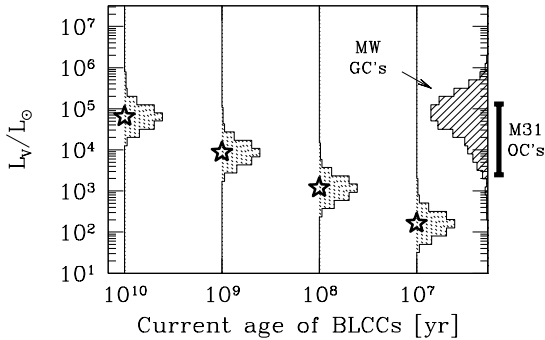,width=\hsize,clip=}}
\caption{The effect of evolution on BLCC luminosity.
The dotted histograms trace the expected BLCCs luminosity function
as predicted at $t = 10$~Gyr, according to different values assumed
for the {\it current} typical age of these objects (as labeled on the
x axis).
We assume a SSP evolution, according to \citet*{buz89} synthesis
models, for a Salpeter IMF and a (roughly) solar metallicity.
For comparison, the observed luminosity distribution of MW GCs is
reported on the right vertical axis, derived from \citet{h96}
(shaded histogram), while the indicative luminosity range for
M31 open clusters is also sketched (thick solid bar on the right)
according to \citet{h_open}.}
\label{lumi}%
\end{figure}

It remains to be explored how BLCCs would appear in the future, e.g., in
particular if they will look like classical globulars when they will become
comparably old.
If we assume BLCCs to consist of plain SSPs, then one should expect  their
luminosity to fade with time, as far as the composing stellar population
becomes older and photometrically dominated by low-mass stars. In particular,
for a SSP of roughly solar metallicity and Salpeter IMF, evolutionary
population  synthesis models predict a quite tuned luminosity change such as
$L_V \propto t^{-0.9}$  over a wide range of age \citep[e.g.][]{tg76,b95a}.
According to the assumed age of present-day BLCCs, then one could infer the
expected luminosity of these clusters at $t = 10$~Gyr and more consistently
compare with the observed luminosity function of old MW GCs. The results of
this illustrative excercise are summarized in Fig.~\ref{lumi}; it is evident
from the figure that, in the more likely case of a current age in the  range
$10^8-10^9$~yrs, BLCCs would end up at $10^{10}$~yrs populating the
low-luminosity  (and low-mass) tail of current MW GC distribution. On the
contrary, in the more extreme (and quite unlikely) case of a current age of 
only  a few $10^7$~yrs
we would be left at 10 Gyr with extremely faint BLCCs, certainly out
of range of the typical MW GCs. Finally, if nowadays BLCCs are already evolved
systems (i.e.\ a few Gyr or older), then  at $t = 10$~Gyr their expected
luminosity will not change  so much and their distribution would maintain them
fully consistent with the bulk of both M31  open clusters and MW GCs. A fair
assessment of the present-day age distribution of this kind of clusters is
therefore a mandatory step to consistently locate them in the appropriate
evolutionary framework. 


\section{Discussion and Summary}

Till a few years ago, the M31 cluster system has been commonly viewed  as an
almost identical (though much larger) analog of the MW GC system, apart from
small differences in average  metallicity (M31 about 0.2 dex richer than MW)
and controversial peculiarities in various spectral indices  \citep[][and
references therein]{bur,burst04}. Recent wide-field imaging  \citep[][and
references therein]{iba04},  together with high precision spectroscopy
\citep{perr_cat,burst04}, have led however to a more detailed recognition and
investigation of M31 substructures, including its GC system, leaving space to
the idea that the Andromeda  and MW cluster systems are actually more different
than conceived so far.  

In particular, \citet{morrison} have put forward a new scenario  suggesting the
existence in M31 of quite sizeable population of GCs associated to the thin
disc and  claimed to be old and metal poor. The natural consequence  of such an
evidence, if confirmed, is that the thin disc was formed when its oldest
metal-poor globulars were in place and that no significant perturbation
affected M31, and in particular its thin disc, since then. An in-depth analysis
of the claimed thin-disc GC members, and in particular  of the metal-poor ones,
may thus greatly help clarifying this important dynamical issue. 

The present study has tackled the problem from a different point of view, i.e.\
by building up a revised global sample of M31 GCs with intrinsically bluer
color than the MW counterparts and assessing if these {\it bona fide} BLCCs
(Sample ``A'' in our analysis) would further discriminate with respect to 
``normal'' M31 and MW globulars as far as other intrinsic properties (like
e.g.\ absolute magnitude, metallicity, spectral indices, structural parameters,
and age) and kinematic behaviour are concerned. On the basis of this
comparison, a tuned and more physical selection  based on the $H\beta$ spectral
index has been proposed (Sample ``B'' in previous discussion). Quite
interestingly, both color- and $H\beta$-selected samples led to fully
consistent conclusions of our analysis.

It has been possible to convincingly  demonstrate here that both 
samples consist of moderately young ($t \lesssim 2$~Gyr) stellar systems, not so
metal poor as previously estimated, and  basically sharing
the thin disc kinematics \citep[see also][on this line]{beas}. Since (a)
essentially all the blue clusters in the thin disc subsystem as defined by 
\citet{morrison} are likely young and (b) the metallicity values adopted by
these authors are sistematically underestimated (at least for most BLCCs, 
see Sect.~3.3), serious doubts are casted on
the actual presence of metal-poor (and old) clusters in the thin disc of M31
as claimed by Morrison et al.\footnote{While the present paper was in the 
peer review phase, a preprint was posted \citep{puzia05}, which also
develops some of the arguments discussed here.}
This argument greatly weakens the possible contrast
with the \citet{bro03} hypothesis of an equal-mass merging event suffered by 
M31 6-8 Gyr ago.

Given the evidences presented so far and within the framework  discussed above,
one might naturally ask whether  also the ``red'' clusters claimed to be member
of the M31 thin disc  subsystem actually are old and metal-poor  (as at least
some of  them seem to appear) or  whether they  could display some spread in
age and/or metal abundance. The answer to such a question is of paramount
importance to settle in finer detail  the mechanisms and formation timescale
for the M31 thin disc and for the global evolution of the galaxy system as a
whole.

\begin{acknowledgements} 
We would like to warmly thank Kathy Perrett for
providing us with her full database of M31 GCs in electronic form. 
This research has been partially supported by the Italian Ministero 
dell'Universit\'a e della Ricerca (MIUR), through the COFIN grant p.\
2002028935-001, assigned to the project  {\it ``Distance and stellar
populations in the galaxies of the Local Group''}.
\end{acknowledgements}

\clearpage

\begin{deluxetable}{lccrcrrcrrrrll}
\tabletypesize{\tiny}
\tablecolumns{14} 
\tablewidth{0pc} 
\tablecaption{The adopted BLCC samples} 
\tablehead{ 
\colhead{Name} &\colhead{V} & \colhead{B-V} & \colhead{U-B} & \colhead{$\langle [Fe/H]\rangle$} &
 \colhead{$\Delta$} & \colhead{H$\beta$} & \colhead{$W_{1/4}$}& \colhead{V$_r$} &\colhead{$\delta$}&\colhead{X} & \colhead{Y} & \colhead{Sample} & \colhead{References}\\
\colhead{} &\colhead{} & \colhead{} & \colhead{} & \colhead{} &
 \colhead{mag} & \colhead{\AA} & \colhead{}&\colhead{km\,s$^{-1}$} & \colhead{}&\colhead{arcmin} & \colhead{arcmin} 
& \colhead{} & \colhead{}}
\startdata 
 B008-G060& 16.56&   1.10&    0.50&    --0.41$\pm$    0.38&  0.397& 3.50&  4.96 &   --319&    0.70& --15.34 &	19.95& B  &		     \\        
 B028-G088& 16.86&   0.88&  --0.05&    --1.87$\pm$    0.29&  0.404& 3.81&  4.80 &   --434&    0.72& --23.65 &	 2.67& B  &  10 	     \\    
 B040-G102& 17.38&   0.29&  --0.01&    --0.98$\pm$    0.48&  0.137& 7.41&  5.32 &   --463&  --0.40& --35.48 & --11.77& A,B&  1,2,6,7	     \\ 
 B043-G106& 16.96&   0.28&  --0.14&    --2.42$\pm$    0.51&  0.144& 5.53&  5.13 &   --414&    0.42& --33.68 & --11.21& A,B&  1,2,3,4,5,6,7,8 \\ 
 B047-G111& 17.51&   0.72&    0.09&    --1.62$\pm$    0.41&\nodata& 3.53&  5.57 &   --291&  --1.02&   13.80 &	24.58& B  &		     \\  
 B049-G112& 17.56&   0.52&    0.18&    --2.14$\pm$    0.55&\nodata& 9.31&  5.83 &   --481&  --0.39& --27.56 &  --7.27& A,B&  7  	     \\   
 B057-G118& 17.64&   0.69& \nodata&    --2.12$\pm$    0.32&  0.289& 5.56&  4.98 &   --437&    0.27& --25.00 &  --7.02& B  &		     \\ 
 B066-G128& 17.42&   0.36&  --0.27&    --2.10$\pm$    0.35&  0.061& 4.67&  4.95 &   --389&    0.70& --29.61 & --13.02& A,B&  3,5,6,7	     \\ 
 B069-G132& 18.16&   0.44&    0.02&    --1.35$\pm$    0.43&  0.338& 7.17&  5.72 &   --295&  --0.27&    3.44 &	11.91& A,B&  7,10	     \\      
 B074-G135& 16.65&   0.75&    0.14&    --1.88$\pm$    0.06&  0.346& 3.92&  5.15 &   --435&  --5.61&   17.38 &	22.06& B  &		     \\     
 B081-G142& 16.80&   0.54&    0.26&    --1.74$\pm$    0.40&  0.352& 7.98&  4.99 &   --430&  --0.47& --25.32 & --12.23& A,B&  7  	     \\   
 B083-G146& 17.09&   0.76&    0.06&    --1.18$\pm$    0.44&  0.047& 3.75&  5.75 &   --367&  --4.15&   19.90 &	22.04& B  &		     \\ 
 B091-G151& 17.56&   0.41&    0.02&    --1.80$\pm$    0.61&  0.269& 7.30&  5.02 &   --290&  --0.16&    2.08 &	 7.01& A,B&  4,7,10	     \\  
 B114-G175& 17.28&   0.42&    0.33&	\nodata	          &\nodata&\nodata&4.90 & \nodata& \nodata&  --3.87 &  --0.58& A  & 10  	     \\ 
 B160-G214& 18.02&   0.55&    0.05&    --1.17$\pm$    1.25&\nodata&\nodata&4.78 &   --354&  --0.46&  --7.98 & --13.45& A  &  10 	     \\
 B170-G221& 17.39&   0.98&    0.53&    --0.54$\pm$    0.24&\nodata& 4.52&  4.63 &   --295&    1.54& --15.59 & --21.43& B  &		     \\  
 B210-M11 & 17.57&   0.52&    0.02&    --1.90$\pm$    0.32&  0.162& 6.83&  4.98 &   --265&    0.11&    7.69 & --12.70& A,B&  7,10,12	     \\ 
 B216-G267& 17.25&   0.20&    0.02&    --1.87$\pm$    0.39&  0.177& 5.66&  5.72 &    --84&  --0.10&   26.91 &	 0.92& A,B&  2,3,5,7,8,10,12 \\   
 B222-G277& 17.43&   0.68&    0.47&    --0.93$\pm$    0.95&  0.349& 8.47&  6.91 &   --311&  --1.07&   10.14 & --16.17& B  & 10,11	     \\  
 B223-G278& 17.81&   0.15&    0.14&    --1.13$\pm$    0.51&  0.297& 4.44&  5.65 &   --101&  --0.01&   26.39 &  --3.81& A,B& 7,10,12	     \\ 
 B237-G299& 17.10&   0.77&    0.16&    --2.09$\pm$    0.28&  0.167& 7.43&  5.08 &    --86&    4.54&   21.81 & --17.48& B  &		     \\ 
 B281-G288& 17.67&   0.84&    0.50&    --0.87$\pm$    0.52&  0.364& 5.56&  6.20 &   --203&    0.76&   16.85 & --15.09& B  &		     \\   
 B295-G014& 16.75&   0.71&    0.27&    --1.71$\pm$    0.15&\nodata& 4.77&  4.92 &   --423& \nodata& --85.95 &	19.69& B  &  8  	     \\ 
 B303-G026& 18.22&   0.24&    0.46&    --2.09$\pm$    0.41&  0.252& 5.78&  5.46 &   --464&    1.84& --65.51 &	 5.53& A,B&  7  	     \\ 
 B307-G030& 17.32&   0.87&    1.01&    --0.41$\pm$    0.36&\nodata& 5.76&  6.35 &   --407&    3.02& --57.96 &	 4.58& B  &		     \\
 B314-G037& 17.63&   0.59&    0.34&    --1.61$\pm$    0.32&  0.402& 5.19&  6.48 &   --485&    0.51& --69.94 & --10.77& B  & 3,8,11	     \\
 B315-G038& 16.47&   0.07&    0.02&    --2.35$\pm$    0.54&  0.116& 4.77&  5.18 &   --559&  --0.18& --55.65 &  --0.83& A,B& 1,2,3,5,7,8,9,12 \\
 B318-G042& 17.02&   0.17&  --0.42&	\nodata	          &\nodata&\nodata&5.31 & \nodata& \nodata& --52.16 &  --1.09& A  & 1,2,3,4,5,7,8    \\ 
 B319-G044& 17.61&   0.72&  --0.64&    --2.27$\pm$    0.47&  0.113& 5.37&  5.49 &   --535&    0.21& --52.03 &  --1.54& B  & 1,2,3,5,7,9      \\   
 B321-G046& 17.67&   0.22&    0.22&    --2.39$\pm$    0.41&  0.229& 6.29&  5.52 &   --518&  --0.05& --55.54 &  --7.17& A,B& 3,7,8,11,12      \\ 
 B322-G049& 17.75&   0.06&  --0.28&	\nodata	          &\nodata&\nodata&5.14 & \nodata& \nodata& --46.31 &  --0.57& A  & 6,7,8,11,12      \\
 B327-G053& 16.58&   0.32&  --0.35&    --2.33$\pm$    0.49&  0.160& 4.09&  4.87 &   --528&    0.25& --47.70 &  --3.23& A,B& 3,4,5,6,7,8,11,12\\ 
 B331-G057& 18.19&   0.25& \nodata&	  \nodata	  &\nodata&\nodata&6.31 & \nodata& \nodata&    4.76 &	36.35& A  &		     \\ 
 B342-G094& 17.73&   0.30&    0.64&    --1.62$\pm$    0.02&  0.003& 7.06&  6.52 &   --479&  --0.66& --40.46 & --12.04& A,B&  1,2,6,7,9       \\
 B354-G186& 17.81&   0.13&    0.69&	\nodata	          &\nodata&\nodata&5.45 & \nodata& \nodata&   35.36 &	26.68& A  &		     \\
 B355-G193& 17.76&   0.53&    0.02&    --1.62$\pm$    0.43&  0.240& 4.39&  4.24 &   --114&    5.25&   34.00 &	24.37& A,B&		     \\
 B358-G219& 15.22&   0.49&    0.19&	 \nodata          &\nodata&\nodata&5.06 & \nodata& \nodata& --64.79 & --58.32& A  &		     \\
 B367-G292& 18.45&   0.32&  --0.17&    --2.32$\pm$    0.53&  0.097& 6.21&  5.64 &   --152&  --0.58&   53.03 &	12.32& A,B&  7  	     \\ 
 B368-G293& 17.92&   0.26&  --0.36&	 \nodata	  &\nodata&\nodata&5.75 & \nodata& \nodata&   41.80 &	 3.36& A  &  1,2,3,5,7,9     \\
 B374-G306& 18.31&   0.44&    0.33&    --1.90$\pm$    0.67&  0.281& 4.07&  5.27 &    --96&    0.88&   41.08 & --10.68& A,B&  7  	     \\
 B376-G309& 18.06&   0.45&    0.23&	  \nodata	  &\nodata&\nodata&5.59 & \nodata& \nodata&   42.12 & --10.79& A  &  3,5,7,8	     \\
 B380-G313& 17.01&   0.47&    0.33&    --2.31$\pm$    0.45&  0.187& 6.52&  6.67 &    --13&    1.14&   58.47 &  --2.07& A,B&  7,8,11,12       \\ 
 B431-G027& 17.73&   0.49&    0.29&	\nodata	          &\nodata&\nodata& 5.59& \nodata& \nodata& --59.11 &	 9.36& A  &  7  	     \\
 B443-D034& 18.20&   0.80&  --0.52&    --2.37$\pm$    0.46&\nodata& 6.72&  5.40 &   --532&  --0.08& --50.47 &  --4.58& B  &		     \\ 
 B448-D035& 17.49&   0.61&    0.01&    --2.16$\pm$    0.19&\nodata& 6.70&  6.54 &   --552&  --0.16& --43.17 &  --2.77& B  &  7  	     \\ 
 B451-D037& 18.66&   0.19& \nodata&    --2.13$\pm$    0.43&\nodata& 3.50&  4.40 &   --514&    0.20& --33.01 &	 2.57& A,B&  7  	     \\
 B453-D042& 17.30&   0.87&    0.16&    --2.09$\pm$    0.53&  0.234& 4.12&  5.27 &   --446&    0.47& --23.69 &	 5.79& B  &		     \\ 
 B458-D049& 17.84&   0.49&    0.91&    --1.18$\pm$    0.67& --0.085& 6.19& 5.98 &   --521&  --0.83& --26.50 &  --6.22& A,B&  7  	     \\ 
 B475-V128& 17.56&   0.31&    0.10&    --2.00$\pm$    0.14&\nodata& 5.96&  7.24 &   --120&  --1.02&   45.00 &	 3.92& A,B&  7  	     \\ 
 B480-V127& 17.91&   0.65&    0.39&    --1.86$\pm$    0.66&  0.188& 5.19&  4.89 &   --135&  --0.50&   44.30 &  --8.38& B  &		     \\ 
 B483-D085& 18.46&   0.27&    0.08&    --2.96$\pm$    0.35&\nodata& 5.58&  5.87 &    --53&  --0.09&   58.16 &	 0.58& A,B&  7  	     \\ 
 B484-G310& 18.10&   0.52&    0.58&    --1.95$\pm$    0.59&  0.110& 5.70&  5.45 &   --104&    0.06&   46.62 &  --8.51& A,B&  7,12	     \\ 
 B486-G316& 17.52&   0.35&    0.93&	 \nodata	  &\nodata&\nodata&5.02 & \nodata& \nodata&    9.45 & --41.39& A  &  7  	     \\
B189D-G047& \nodata& \nodata& \nodata& --1.19$\pm$    0.29& --0.064& 4.24& \nodata&   --584&--0.62& --45.82 &	 0.24& B  &		     \\
      VDB0& 15.28&   0.23&  --0.37&	 \nodata	  &\nodata&\nodata&\nodata& \nodata&\nodata&--47.41 &  --4.31& A  &  3,4,5,6	     \\ 
  NB21-AU5& 17.86&   0.31& \nodata&	 \nodata          &\nodata&\nodata&\nodata& \nodata&\nodata& --0.88 &	 0.85& A  &		     \\
 NB67-AU13& 16.14&   0.48&  --0.03&    --1.43$\pm$    0.13&\nodata&\nodata&\nodata&   --113&  1.66&    1.69 &	 3.75& A  &  7  	     \\
      NB83& 16.68&   0.56&  --0.03&    --1.26$\pm$    0.16&\nodata&\nodata&\nodata&   --150&  1.30&  --4.24 &	 0.89& A  &		     \\
B006D-D036& 18.00& \nodata& \nodata&    --2.16$\pm$   0.32& --0.026& 5.04&\nodata&   --522&   0.51& --36.31 &	 2.25& B  &		     \\
B012D-D039& 18.40& \nodata& \nodata&    --1.22$\pm$   0.41&  0.182& 7.09&\nodata&   --478&  --0.10& --26.72 &	 6.07& B  &		     \\
B015D-D041& 17.80& \nodata& \nodata&    --1.14$\pm$   0.30&\nodata& 7.32&\nodata&   --445&  --0.41& --19.24 &	 9.37& B  &		     \\ 
B111D-D065& 18.10& \nodata& \nodata&    --1.80$\pm$   0.36&  0.082& 5.55&\nodata&   --130&  --0.81&   27.38 &	 2.27& B  &		     \\ 
     B195D& 15.19&   0.22&  --0.30&    --1.64$\pm$    0.19&  0.166& 4.29&\nodata&   --552&  --0.45& --47.19 &  --4.17& A,B&		     \\  
B206D-D048& 19.06&   0.12& \nodata&    --2.01$\pm$    0.99&\nodata& 2.53&\nodata&   --490&  --0.07& --27.97 &  --6.40& A  &		     \\
B257D-D073& 17.40& \nodata& \nodata&    --1.99$\pm$   0.19&  0.501& 5.49&\nodata&   --114&  --1.25&   46.00 &	 3.87& B  &		     \\   
     DAO47& \nodata& \nodata& \nodata&  --1.13$\pm$   0.57&  0.229& 4.03&\nodata&   --490&  --0.35& --33.09 &  --7.67& B  &		     \\
      V031& 17.43&   0.68&    0.39&    --1.59$\pm$    0.06&  0.497& 5.84&\nodata&   --433&    0.02& --19.01 &	 7.27& B  &  12 	     \\
\enddata
\label{tab1}
\tablerefs{ (1) \cite{syd67}; (2) \cite{sg77}; (3) \cite{EW88}; (4) \cite{boh88}; 
(5) \cite{kl91}; (6) \cite{bohlin}; (7) \cite{barmby}; (8) \cite{barm01}; 
(9) \cite{wh01}; (10) \cite{Ji03}, (11) \cite{beas}; (12) \cite{burst04}}
\end{deluxetable} 										     

\begin{deluxetable}{lccrcrrcrrrrl}
\tabletypesize{\scriptsize}
\tablecolumns{13} 
\tablewidth{0pc} 
\tablecaption{Other confirmed and reportedly ``young'' M31 GCs} 
\tablehead{ 
\colhead{Name} &\colhead{V} & \colhead{B-V} & \colhead{U-B} & \colhead{$\langle [Fe/H]\rangle$} &
 \colhead{$\Delta$} & \colhead{H$\beta$} & \colhead{$W_{1/4}$}&\colhead{V$_r$} &  \colhead{$\delta$}&\colhead{X} & \colhead{Y} & \colhead{References}\\
\colhead{} &\colhead{} & \colhead{} & \colhead{} & \colhead{} &
 \colhead{mag} & \colhead{\AA} & \colhead{}&\colhead{km\,s$^{-1}$} & \colhead{}&\colhead{arcmin} & \colhead{arcmin} 
& \colhead{}}
\startdata 
    B015-V204& 17.79&	1.41& \nodata& --0.35$\pm$  0.96& \nodata& --0.54&   5.56&   --460&  --0.04 & --26.56 &    7.91&  10	  \\ 
    B030-G091& 17.39&	1.93&    0.71& --0.39$\pm$  0.36& \nodata&   1.62&   4.69&   --380&    1.62 & --24.83 &    1.24&  10	  \\
    B090     & 18.80&\nodata& \nodata& --1.39$\pm$  0.80& \nodata&   3.28&   4.62&   --428&  --0.43 & --13.12 &  --4.60&  10	  \\ 
    B101-G164& 16.87&	0.81&    0.38&   \nodata	& \nodata&\nodata&   4.49& \nodata& \nodata &  --8.01 &  --2.55&  10	  \\ 
 	 B102& 16.58&	0.62&  --0.12& --1.57$\pm$  0.10&   0.224&   2.74&   3.98&   --236&    0.17 &	12.67 &   13.29&   7	  \\ 
    B117-G176& 16.34&	0.65&    0.45& --1.33$\pm$  0.45&   0.266&   2.70&   4.76&   --531&  --2.68 & --16.19 & --10.13&   8	  \\ 
    B146     & 16.95&	1.49&  --0.54&    \nodata	& \nodata&\nodata&   4.86& \nodata& \nodata &	 1.51 &  --3.23&  10	  \\ 
    B154-G208& 16.82&	1.32&    0.55&   \nodata	& \nodata&\nodata&   4.95& \nodata& \nodata &	 3.17 &  --4.21&  10	  \\ 
    B164-V253& 17.94&	1.04& \nodata& --0.09$\pm$  0.40&   0.559&   1.48&   4.74&   --294&    0.07 &	 1.00 &  --7.25&  10	  \\
    B197-G247& 17.63&	1.08&    0.19& --0.43$\pm$  0.36& \nodata&   1.14&   4.94&     --9&    1.30 &	18.58 & --1.06 &  10	  \\ 
    B214-G265& 17.65&	0.61&    0.28& --1.00$\pm$  0.61& \nodata&   3.24&   4.90&   --258&  --1.26 &	17.19 &  --5.53&   7	  \\ 
    B232-G286& 15.67&	0.72&    0.10& --1.83$\pm$  0.14&   0.242&   3.13&   4.73&   --179&    2.52 &	12.52 & --17.88&  12	  \\ 
    B292-G010& 16.99&	0.90&  --0.02&      \nodata	& \nodata&\nodata&   5.30& \nodata& \nodata & --58.32 &   47.44&  11	  \\ 
    B311-G033& 15.44&	0.96&    0.14& --1.96$\pm$  0.07&   0.120&   2.72&   4.53&   --463&    2.15 & --57.58 &    1.24&  12	  \\
B324-G051$^a$& 16.91&   0.66&    0.79&      \nodata	& \nodata&\nodata&   4.87& \nodata& \nodata &	 3.10 &   36.44&   8,11,12\\
    B328-G054& 17.86&	0.89&    0.48&      \nodata	& \nodata&\nodata&   5.44& \nodata& \nodata &	 3.20 &   35.57&  7,10    \\
    B347-G154& 16.50&	0.73&    0.15&      \nodata	& \nodata&\nodata&   5.16& \nodata& \nodata &	27.83 &   26.66&  8,12    \\
 	 B423& 17.72&	0.60&    0.10&      \nodata	& \nodata&\nodata&   6.33& \nodata& \nodata & --47.66 &   31.80&   7	  \\  
 	 B468& 17.79&	0.70& \nodata&      \nodata	& \nodata&\nodata&   7.52& \nodata& \nodata & --66.43 & --58.30&   7	  \\
 	 NB16& 17.55&	0.66& \nodata& --1.36$\pm$  0.12& \nodata&\nodata&\nodata&   --115&    1.35 &	 1.97 &    4.21&   8	  \\
 	B150D& 17.55&	0.61&  --0.04&      \nodata	& \nodata&\nodata&\nodata& \nodata& \nodata & --31.77 &  57.17&   7	  \\
\enddata 
\label{tab2}
\tablecomments{$^a$ Photometry from \cite{sha95}}
\tablerefs{ 
(1) \cite{syd67}; (2) \cite{sg77}; (3) \cite{EW88}; (4) \cite{boh88}; 
(5) \cite{kl91}; (6) \cite{bohlin}; (7) \cite{barmby}; (8) \cite{barm01}; 
(9) \cite{wh01}; (10) \cite{Ji03}, (11) \cite{beas}; (12) \cite{burst04}}
\end{deluxetable} 
	
\begin{deluxetable}{lcrrrrl}
\tabletypesize{\scriptsize}
\tablecolumns{7} 
\tablewidth{0pc} 
\tablecaption{Unconfirmed M31 GCs, reportedly ``young'' and/or\protect \\
 possible BLCCs candidates with $(B-V)_o<0.45$} 
\tablehead{ 
\colhead{Name} & \colhead{V} & \colhead{B-V} & \colhead{U-B}  &
\colhead{X} & \colhead{Y} & \colhead{References}\\
\colhead{} &\colhead{} & \colhead{} & \colhead{} & \colhead{arcmin} & \colhead{arcmin} &
\colhead{}}
\startdata 
  B060-G121& 16.75 &   0.71&   0.08 &  --14.10&    0.38  & 7   \\ 
  B070-G133& 17.07 &   0.54&   0.19 &  --10.83&    0.56  &     \\ 
       B089& 18.18 &   0.10& --0.32 &    16.05&   17.93  & 7   \\ 
  B100-G163& 17.91 &   0.88&   0.15 &  --22.53& --13.75  & 7   \\ 
  B108-G167& 17.47 &   0.89&   1.01 &   --7.30&  --2.49  & 7,10\\ 
       B145& 18.10 &   0.32&   1.48 &   --0.95&  --4.82  & 6,7 \\ 
  B150-G203& 16.80 &   1.10&	0.39&     5.97&  --0.87  & 10  \\			     
  B157-G212& 17.73 &   0.65& --0.11 &   --0.39&  --7.36  & 7   \\ 
  B173-G224& 18.27 &   0.02& \nodata&    10.21&  --2.61  &     \\ 
  B192-G242& 18.28 &   0.20&   0.32 &    23.73&    4.21  & 7,10\\ 
       B195& 18.57 &   0.40&   0.97 &   --3.39& --17.95  & 7   \\ 
       B323& 17.59 &   0.47&   0.17 &  --51.25&  --4.57  &     \\ 
  B330-G056& 17.72 &   0.95&   0.09 &     5.33&   36.93  & 3   \\ 
  B362     & 17.61 &   0.65&   0.08 &    29.90&    3.14	 & 10  \\				     
  B371-G303& 17.54 &   0.48&   0.39 &    40.53&  --7.04  & 7   \\ 
       B414& 17.98 &   0.50&   0.08 &     4.58&   93.72  &     \\ 
  B442-D033& 17.94 &   0.39&   0.16 &  --47.36&  --1.87  & 7   \\ 
  B452-G069& 17.78 &   0.38& --0.05 &  --45.89&  --7.66  & 7   \\ 
       B460& 18.35 &   0.54& --0.23 &  --85.52& --54.13  &     \\ 
  B469-G220& 17.58 &   0.53& --0.04 &    46.76&   28.06  & 3,7 \\ 
  B477-D075& 18.46 &   0.32&\nodata &    35.11&  --6.78  &     \\ 
       B508& 17.12 &   0.46&   0.27 &    54.77& --57.54  &     \\ 
 B190D-G048& 18.17 &   0.24& --0.29 &  --46.03&  --0.26  &     \\ 
  G085-V015& 17.39 &   0.23&   0.05 &  --43.73& --11.86  &     \\ 
 B028D-G100& 18.04 &   0.40& --0.21 &  --26.71&  --4.77  &     \\ 
       G137& 17.81 & --0.02& --1.16 &     5.75&   12.76  &     \\ 
       G270& 17.30 &   0.46& --0.77 &    23.70&  --2.39  &     \\ 
       H126& 16.76 &   0.42& --0.03 &    43.75&   13.32  &     \\ 
   NB39-AU6& 17.94 &   0.21&   0.40 &     0.18&  --0.79  &     \\ 
       NB42& 18.49 &   0.52& \nodata&     1.49&    0.56  &     \\ 
   NB47-AU3& 18.75 &   0.10& \nodata&     1.44&    3.59  &     \\ 
 B065D-NB69& 16.83 &   0.37& \nodata&     1.41&    2.52  &     \\ 
       NB79& 18.27 &   0.53& \nodata&   --2.87&    2.01  &     \\ 
      NB107& 18.65 &   0.31& \nodata&   --2.28&    1.62  &     \\ 
      B134D& 18.19 &   0.46& --0.19 &  --74.79&   46.10  &     \\ 
      B137D& 18.51 &   0.54& --0.05 &  --59.80&   45.32  &     \\ 
      B139D& 18.59 &   0.44& \nodata& --116.18&    2.35  &     \\ 
      B147D& 17.96 &   0.49& --0.44 &  --47.16&   49.24  &     \\ 
      B158D& 16.50 &   0.33&   0.02 & --103.80&  --5.69  &     \\ 
      B162D& 17.85 &   0.44& --0.52 &  --91.16&  --0.17  &     \\ 
      B171D& 17.80 &   0.43&   0.08 &  --57.63&   14.65  &     \\ 
      B173D& 17.45 &   0.55& --0.43 &  --54.91&   15.16  &     \\ 
      B196D& 18.79 & --0.21& --0.94 &  --54.20& --10.79  &     \\ 
      B216D& 18.21 &   0.30&   0.01 &    28.36&   21.50  &     \\ 
      B218D& 18.19 &   0.44& --0.02 &    41.70&   28.85  &     \\ 
      B220D& 16.94 &   0.53& --0.06 &  --65.70& --55.73  &     \\ 
      B225D& 18.15 &   0.50& --0.19 &  --55.82& --50.27  &     \\ 
      B227D& 16.91 & --0.01&   0.03 &  --44.15& --41.65  &     \\ 
      B246D& 18.05 &   0.09& --0.77 &    49.50&   15.14  &     \\ 
      B253D& 17.68 &   0.53&   0.03 &  --16.92& --42.36  &     \\ 
      B261D& 17.60 &   0.46& --0.04 &    60.74&   11.30  &     \\ 
      B270D& 17.50 &   0.40&   0.34 &    10.14& --36.21  &     \\ 
      B272D& 16.34 &   0.51&   0.37 &  --24.27& --64.72  &     \\ 
      B286D& 17.76 &   0.47& --0.91 &   --7.48& --59.92  &     \\ 
      B293D& 17.91 &   0.54& --0.12 &  --29.58& --81.90  &     \\ 
      B303D& 18.27 &   0.31&   0.91 &  --43.52& --97.03  &     \\ 
      B312D& 18.35 &   0.53& \nodata&  --24.72& --88.18  &     \\ 
      B320D& 17.78 &   0.45& --0.56 &    72.08& --13.85  &     \\ 
      B322D& 18.26 &   0.17&   0.29 &  --36.49&--101.57  &     \\ 
      B324D& 18.74 &   0.55& --0.03 &    58.96& --27.38  &     \\ 
  DA046$^a$& 18.75 &   0.53& \nodata&  --47.09& --16.89  & 7   \\ 	
      DA052& 18.42 &   0.14& --0.50 &  --24.11& --10.15  & 7   \\ 
  DA069$^a$& 17.48 &   0.26& \nodata&    42.11&    6.66  & 7   \\	
       V014& 17.39 &   0.35&   0.04 &  --43.86& --12.20  & 1   \\ 
       V034& 17.48 &   0.13& --1.15 &  --17.30&    4.24  &     \\ 
       V133& 18.36 &   0.06& --1.26 &    51.54&    5.58  &     \\ 
       V270& 18.07 &   0.27& --0.67 &    12.26&    9.93  &     \\ 
       SH06& 16.54 &   0.28& --0.71 &  --27.99&   27.16  &     \\ 
       BH02& 18.50 &   0.40& \nodata&  --49.25&  --1.15  &     \\ 
\enddata 
\label{tab3}
\tablecomments{$^a$ Photometry from \cite{barmby}}
\tablerefs{
(1) \cite{syd67}; (2) \cite{sg77}; (3) \cite{EW88}; (4) \cite{boh88}; 
(5) \cite{kl91}; (6) \cite{bohlin}; (7) \cite{barmby}; (8) \cite{barm01}; 
(9) \cite{wh01}; (10) \cite{Ji03}, (11) \cite{beas}; (12) \cite{burst04}}
\end{deluxetable} 										     

\end{document}